\documentclass[aps,prc,twocolumn,10pt, superscriptaddress, showpacs, floatfix]{revtex4-1}

\usepackage{amssymb,epsfig}
\usepackage{bbold}

\usepackage[dvipsnames]{xcolor}

\newcommand{\be}{\begin{equation}}
\newcommand{\ee}{\end{equation}}
\newcommand{\bea}{\begin{eqnarray}}
\newcommand{\eea}{\end{eqnarray}}

\begin{document}

\title{Response of superfluid fermions at finite temperature}
%\title{Finite-temperature microscopic response theory for strongly-coupled superfluid fermionic systems}

\author{Sumit Bhattacharjee}
\affiliation{Department of Physics, Western Michigan University, Kalamazoo, MI 49008, USA}
\author{Elena Litvinova} 
\affiliation{Department of Physics, Western Michigan University, Kalamazoo, MI 49008, USA}
%\affiliation{National Superconducting Cyclotron Laboratory, Michigan State University, East Lansing, MI 48824, USA}
%\affiliation{GANIL, CEA/DRF-CNRS/IN2P3, F-14076 Caen, France}
\date{\today}

\begin{abstract}
A consistent finite-temperature microscopic theory for the response of strongly coupled superfluid fermionic systems is formulated. 
We start from the general many-body Hamiltonian with the vacuum (bare) two-fermion interaction and derive the equation of motion (EOM) for the thermally averaged two-time two-fermion correlation function, which determines the spectrum of the system under study. The superfluidity is introduced via the Bogoliubov transformation of the fermionic field operators, and the entire formalism is carried out in the basis of Bogoliubov's quasiparticles, keeping the complete $4\times 4$ block matrix structure of the two-fermion EOM. Fully correlated static and dynamical interaction kernels of the resulting EOM are discussed. 
%The least correlated static kernel reproduces the known finite-temperature quasiparticle random phase approximation, while the dynamical kernel was previously largely unknown in the finite-temperature superfluid theory. 
A special focus is then placed on the latter kernel, which is advanced to a factorized form, enabling a minimal truncation of the many-body problem while keeping important effects of emergent collectivity and mapping to the quasiparticle-vibration coupling (qPVC). As in the zero-temperature and non-superfluid cases, the qPVC can be associated with a new order parameter, qPVC vertex. In the thermal superfluid theory, the latter vertex as well as the components of the dynamical kernel acquire an extended form including thermally unblocked transition amplitudes.
%.  allows for building approximations of growing complexity.
%takes over the power countings dominating the bare fermionic interaction in systems with pronounced collectivity. 
%Finally, the dynamical qPVC kernel is brought to the form enabling numerical implementations for realistic strongly coupled fermionic systems, in particular, %medium-mass and heavy nuclei.  

\end{abstract}
%\pacs{21.10.-k, 21.30.Fe, 21.60.-n, 23.40.-s, 24.10.Cn, 24.30.Cz}

\maketitle

%===============================================================================
% Introduction
%===============================================================================
\section{Introduction} 

The response of interacting many-body fermionic systems to external probes is one of their most informative characteristics. The response functions are particularly sensitive to microscopic details of the interparticle interaction and its modification in correlated media and hold correspondence with the spatial arrangement of quantum matter.  They, therefore, constitute an efficient formalism for quantifying emergent properties, such as collective excitation modes, superfluidity, and superconductivity.

%{\color{ForestGreen} 2$^-$}

Response functions can be conveniently related to correlation functions (CFs) of the time-dependent fermionic field operators and play the role of scattering amplitudes in correlated media at zero and finite temperatures. 
Such CFs form the common formal quantum field theory (QFT) background across the energy scales, from particle physics to quantum chemistry. 
Many-body systems with medium and strong coupling require non-perturbative approaches, in contrast to weakly coupled ones, which can be reasonably tackled by perturbation theory with a small parameter, typically associated with the interaction strength. A non-perturbative treatment can be organized by rearranging the many-body Hamiltonian in a way the perturbation theory can still be applied or, alternatively, by different criteria. 
% EFT is derivable, place this below
In both cases, an adequate quantification of microscopic mechanisms of emergent collectivity and its thermal evolution plays an important role in solving the quantum many-body problem at different scales.

The CFs in nuclear matter and atomic nuclei are determined by intermediate coupling and complicated by the uncertainties of the nucleon-nucleon (NN) interaction. Moreover, the nuclear response has a complex multichannel structure tangled by the in-medium dynamics. 
%However, the same feature stipulates the relevance of a large variety of experimental probes, which are sensitive to certain channels and, thus, can illuminate particular aspects of the strong %These opportunities are actively explored by the rare isotope beam facilities \cite{Tanihata1998,Glasmacher2017}.  
Nuclear response theory pursues a microscopic approach to describe nuclear spectral properties, that is, excited states and transition probabilities, as accurately as possible. The gross and fine features of the nuclear strength functions, which quantify the observable features of the nuclear response, are of prime importance and provide critical input to numerous nuclear science applications. For instance, astrophysical modeling of the r-process nucleosynthesis is largely based on the nuclear strength functions of the dipole character, in particular, electric dipole (E1), magnetic dipole (M1), Gamow-Teller (GT$_-$), and spin-dipole (SD) ones, which determine radiative neutron capture and beta decay rates. Electron capture rates in the core-collapse supernovae are extracted from the GT$_+$ strength distributions. The QFT formalism adopted for the non-perturbative physics of the nuclear medium enables the formulation of a generic response theory applicable to all these phenomena at both zero and finite temperatures. The latter becomes a relevant factor for nuclear structure in hot stellar environments, which modifies the low-energy spectra and associated reaction rates considerably. 

It is recognized that the nuclear response theory capable of quantitatively addressing nuclear spectra must contain in-medium induced NN interaction, which is beyond the reach of the random phase approximation (RPA) \cite{Bohm1951}, or its superfluid variant, the quasiparticle RPA (QRPA) \cite{Baranger1960,RingSchuck1980}. 
The leading effects are attributed to the minimal coupling between the single-particle and emergent collective degrees of freedom \cite{BohrMottelson1969,BohrMottelson1975,Broglia1976,BortignonBrogliaBesEtAl1977,BertschBortignonBroglia1983,Soloviev1992} which cause fragmentation of the (Q)RPA states or damping of collective excitations. On the formal level, the quasiparticle-vibration coupling (qPVC) can be linked to the dynamical kernels of the equations of motion (EOM) for the CFs in nuclear medium \cite{RingSchuck1980,Schuck1976,AdachiSchuck1989,Danielewicz1994,DukelskyRoepkeSchuck1998,SchuckTohyama2016,LitvinovaSchuck2019,LitvinovaSchuck2020}. In the regime of the NN intermediate coupling, qPVC emerges with a new order parameter associated with the qPVC vertex, which can be used to produce a hierarchy of approximations to nucleonic
propagators. The leading-order qPVC kernels can be mapped to the kernels of nuclear field theory (NFT) and related to the formalism of the quasiparticle-phonon models (QPM) \cite{BohrMottelson1969,BohrMottelson1975,Broglia1976,BortignonBrogliaBesEtAl1977,BertschBortignonBroglia1983,Tselyaev1989,KamerdzhievTertychnyiTselyaev1997,Tselyaev2013,Soloviev1992} in the fully microscopic EOM framework \cite{LitvinovaSchuck2019,LitvinovaSchuck2020}. The latter allowed for establishing a connection between the ab initio and effective theories and extending NFT to more complex configurations.

In Ref. \cite{Litvinova2022}, the two-fermion EOM approach was advanced to the superfluid case by working out the complete formalism in the space of the Bogoliubov quasiparticles and keeping the 2$\times$2 matrix structure of the Hartree-Fock-Bogoliubov (HFB) basis. As in the non-superfluid phase, the modification of the bare interaction in the strongly correlated medium was derived, revealing the splitting of the interaction kernel into the static and dynamical components, in principle, calculable from the underlying bare interactions.  The static kernel is associated with short-range correlations, while its dynamical counterpart generates long-range correlations originating from the retardation effects of the correlated medium, and these correlations, in turn, affect the static kernel. The major theoretical step made in Ref. \cite{Litvinova2022} was the unification of the particle-hole ($ph$) and particle-particle ($pp$) channels in one EOM with the interaction kernel unifying the normal and pairing phonons and thus extending the applicability of the theory to open-shell nuclei, which constitute the major part of the nuclear landscape. 

The finite-temperature variants of the response theory beyond QRPA were worked out for the $ph$ and $pp$ channels separately in Refs. \cite{LitvinovaWibowo2018,WibowoLitvinova2019,LitvinovaRobinWibowo2020} and \cite{Litvinova2021}, respectively. Applications to the electric dipole strength \cite{LitvinovaWibowo2018,WibowoLitvinova2019}, beta decay \cite{LitvinovaRobinWibowo2020}, and electron capture rates \cite{Litvinova2021b} revealed the necessity of completion of the theory by developing a finite-temperature formalism including superfluidity in a unified way. This is done in the present work coherently. We derive and discuss the most general EOM and leading approximations to the interaction kernel. We benchmark the static kernel of the superfluid response by the finite-temperature QRPA \cite{Sommermann1983,RingRobledoEgidoEtAl1984} and then focus on the qPVC variants of the dynamical kernel, bringing them to the form ready for numerical implementations, which will be presented in a separate article.

%===============================================================================
% Formalism
%===============================================================================
\section{Theory}
\label{Propagators}
\subsection{Response to an external field in the quasiparticle formalism: the generic EOM}

\subsubsection{Zero temperature}

The minimal input to the many-body theory is the fermionic Hamiltonian $H$ 
\be
H = H^{(1)} + V^{(2)},
\label{Hamiltonian}
\ee
consisting of the one-body and two-body parts $H^{(1)}$ and $V^{(2)}$, respectively. More complex terms will not be considered in this work but can be included straightforwardly following the same logic. In the second quantized form, the one-body part $H^{(1)}$ reads
\be
H^{(1)} = \sum_{12} t_{12} \psi^{\dag}_1\psi_2 + \sum_{12}v^{(MF)}_{12}\psi^{\dag}_1\psi_2 \equiv \sum_{12}h_{12}\psi^{\dag}_1\psi_2,
\label{Hamiltonian1}
\ee
in terms of the fermionic field operators $\psi_1$ and $\psi^{\dagger}_1$ satisfying the usual anticommutation relations. The matrix elements $h_{12}$ combine the kinetic energy $t$ and external mean field $v^{(MF)}$, in case the latter is present. The number subscripts stand for a working single-particle basis, which we do not specify until explicitly stated. The two-fermion interaction $V^{(2)}$, in this setting, has the form  
\be
V^{(2)} = \frac{1}{4}\sum\limits_{1234}{\bar v}_{1234}\psi^{\dagger}_1\psi^{\dagger}_2\psi_4\psi_3,
\label{Hamiltonian2}
\ee
with ${\bar v}_{1234} = v_{1234} - v_{1243}$ being the antisymmetrized matrix elements of the interaction between two fermions in free space. For a self-bound many-body system, $V$ in the analytical form or in the form of the two-body matrix elements should define the properties of the system, in principle, uniquely. Here, we assume only the instantaneous character of the interaction without detailing its other properties. This assumption should be reasonable enough for the low-energy fermionic systems of interest. 
% Put a note about convenience of the quasiparticle basis, while in the canonical basis there would be 16 propagators to calculate.
A superfluid system of fermions is most conveniently described in the space of Bogolyubov's quasiparticles, which mix particle and hole states via the transformation \cite{Bogolyubov1958,RingSchuck1980}
\bea
\psi_1 = \sum\limits_{\mu} \bigl(U_{1\mu}\alpha_{\mu} + V^{\ast}_{1\mu}\alpha^{\dagger}_{\mu}\bigr) \nonumber\\
\psi^{\dagger}_1 =  \sum\limits_{\mu} \bigl(V_{1\mu}\alpha_{\mu} + U^{\ast}_{1\mu}\alpha^{\dagger}_{\mu}\bigr),
\label{Btrans}
\eea
 or, in the operator form:
\bea
\left( \begin{array}{c} \psi \\ \psi^{\dagger} \end{array} \right) = \cal{W} \left( \begin{array}{c} \alpha \\ \alpha^{\dagger} \end{array} \right), \ \ \ \ \ \ \ \
\cal{W} = \left( \begin{array}{cc} U & V^{\ast} \\ V & U^{\ast} \end{array} \right).
\label{W}
\eea
%where
%\bea
%\cal{W} = \left( \begin{array}{cc} U & V^{\ast} \\ V & U^{\ast} \end{array} \right) \ \ \ \ \ \  \cal{W}^{\dagger} = \left( \begin{array}{cc} U^{\dagger} & V^{\dagger} \\ V^T & U^T \end{array} \right). 
%\eea
In Eqs. (\ref{Btrans},\ref{W}) and henceforth, the Greek subscripts will be used to denote the basis of quasiparticle states. 
The transformation $\cal W$ is unitary as the quasiparticle operators $\alpha$ and $\alpha^{\dagger}$ form the same anticommutator algebra as the particle operators.
The matrices $U$ and $V$ are determined from the ground state energy solution, e.g., minimization on some many-body wave function ansatz. One common approach is the HFB one. The $U$ and $V$ matrices satisfy the following requirements:
\bea
U^{\dagger}U + V^{\dagger}V = \mathbb{1}\ \ \ \ \ \ UU^{\dagger} + V^{\ast}V^{T} = \mathbb{1}\nonumber\\
U^TV + V^TU = 0\ \ \ \ \ \  UV^{\dagger} + V^{\ast}U^{T} = 0 .
\label{UV}
\eea

The Hamiltonian (\ref{Hamiltonian}) is transformed to the quasiparticle basis (\ref{Btrans}) by applying the transformation $\cal W$ to the fermionic operators in Eq. (\ref{Hamiltonian}).%,  takes the form \cite{RingSchuck1980}:
%\bea
%H = H^0 &+& \sum\limits_{\mu\nu}H^{11}_{\mu\nu}\alpha^{\dagger}_{\mu}\alpha_{\nu}  + \frac{1}{2}\sum\limits_{\mu\nu}\bigl(H^{20}_{\mu\nu}\alpha^{\dagger}_{\mu}\alpha^{\dagger}_{\nu} + %\text{h.c.}\bigr) \nonumber\\
%&+& \sum\limits_{\mu\mu'\nu\nu'}\bigl(H^{40}_{\mu\mu'\nu\nu'}\alpha^{\dagger}_{\mu}\alpha^{\dagger}_{\mu'}\alpha^{\dagger}_{\nu}\alpha^{\dagger}_{\nu'} + \text{h.c}\bigr) \nonumber\\
%&+& \sum\limits_{\mu\mu'\nu\nu'}\bigl(H^{31}_{\mu\mu'\nu\nu'}\alpha^{\dagger}_{\mu}\alpha^{\dagger}_{\mu'}\alpha^{\dagger}_{\nu}\alpha_{\nu'} + \text{h.c}\bigr) 
%\nonumber\\
%&+& \frac{1}{4}\sum\limits_{\mu\mu'\nu\nu'}
%\bigl(
%H^{22}_{\mu\mu'\nu\nu'}\alpha^{\dagger}_{\mu}\alpha^{\dagger}_{\mu'}\alpha_{\nu'}\alpha_{\nu}% + \text{h.c}\bigr), 
%\label{Hqua}
%\eea
The resulting form of it reads
 \cite{RingSchuck1980}:
 \be
 H = H^0 + \sum\limits_{\mu}E_{\mu}\alpha^{\dagger}_{\mu}\alpha_{\mu} + V,
 \label{Hqua1}
 \ee
which regroups the contributions according to the number of quasiparticle operators into the non-operator $H^0$, one-quasiparticle term $\alpha^{\dagger}_{\mu}\alpha_{\mu}$ with coefficients $H^{11}$ corresponding to the quasiparticle energies, that is $H^{11}_{\mu\nu} = \delta_{\mu\nu}E_{\mu}$, and the residual interaction $V$ absorbing all the contributions with four quasiparticle operators:
 \bea
 V &=& 
\sum\limits_{\mu\mu'\nu\nu'}\bigl(H^{40}_{\mu\mu'\nu\nu'}\alpha^{\dagger}_{\mu}\alpha^{\dagger}_{\mu'}\alpha^{\dagger}_{\nu}\alpha^{\dagger}_{\nu'} + \text{h.c}\bigr) \nonumber\\
&+& \sum\limits_{\mu\mu'\nu\nu'}\bigl(H^{31}_{\mu\mu'\nu\nu'}\alpha^{\dagger}_{\mu}\alpha^{\dagger}_{\mu'}\alpha^{\dagger}_{\nu}\alpha_{\nu'} + \text{h.c}\bigr) 
\nonumber\\
&+& \frac{1}{4}\sum\limits_{\mu\mu'\nu\nu'}
%\bigl(
H^{22}_{\mu\mu'\nu\nu'}\alpha^{\dagger}_{\mu}\alpha^{\dagger}_{\mu'}\alpha_{\nu'}\alpha_{\nu}.% + \text{h.c}\bigr).
\label{Hqua2} 
\eea 
The upper indices in the matrix elements $H^{ij}_{\mu\nu\mu'\nu'}$ are associated with the numbers of creation and annihilation quasiparticle operators, respectively. The explicit form of the matrix elements $H^{ij}_{\mu\nu\mu'\nu'}$ can be found, e.g., in Ref. \cite{RingSchuck1980}. The $H^{20}$ term vanishes at the stationary point characterizing the HFB ground state, so it will be omitted in this work.

%\subsection{Strength function and superfluid response}

In this paragraph, we are interested in the energies of excited states and probabilities of transitions from the ground to these excited states. Both characteristics can be conveniently described by the strength function associated with a particular excitation operator of an external field $F$. For reasonably weak fields, the strength function $S(\omega)$ reads:
\be
S(\omega) = \sum\limits_{n>0} \Bigl[ |\langle n|F^{\dagger}|0\rangle |^2\delta(\omega-\omega_n) - |\langle n|F|0\rangle |^2\delta(\omega+\omega_n)
\Bigr],
\label{SF}
\ee
where $\omega$ is the transition frequency (energy), and the summation runs over all the formally exact excited states $|n\rangle$. The transition probabilities of absorption and emission, respectively, are
\be
B_n = |\langle n|F^{\dagger}|0\rangle |^2\ \ \ \ \ \ \ \ 
{\bar B}_n = |\langle n|F|0\rangle |^2.
\label{Prob}
\ee
In this work, we consider operators of one-body character $F = \sum_{12}f_{12}\psi_1^{\dagger}\psi_2$, with the quasiparticle operator expansion 
\bea
F = \frac{1}{2}\sum\limits_{\mu\mu'} \Bigl(F^{02}_{\mu\mu'}\alpha_{\mu'}\alpha_{\mu} 
+ F^{20}_{\mu\mu'}\alpha^{\dagger}_{\mu}\alpha^{\dagger}_{\mu'}  
+ F^{11}_{\mu\mu'}\alpha^{\dagger}_{\mu}\alpha_{\mu'}  \nonumber \\
+ {\tilde F}^{11}_{\mu\mu'}\alpha^{\dagger}_{\mu'}\alpha_{\mu} \Bigr) + \sum_{\mu}F^{00}_{\mu}, \ \ \ \ \ \ \ 
\nonumber\\
%F^{\dagger} = \frac{1}{2}\sum\limits_{\mu\mu'} \Bigl(F^{20\ast}_{\mu\mu'}\alpha_{\mu'}\alpha_{\mu} +
%F^{02\ast}_{\mu\mu'}\alpha^{\dagger}_{\mu}\alpha^{\dagger}_{\mu'}  
%\Bigr),
\label{Fext}
\eea
where
\bea
F^{02}_{\mu\mu'} &=& -\sum\limits_{12}V^T_{\mu 1}f_{12}U_{2\mu'} \ \ \ \ \ \ \ F^{20}_{\mu\mu'} = \sum\limits_{12}U^{\dagger}_{\mu 1}f_{12}V^{\ast}_{2\mu'} \nonumber\\
F^{11}_{\mu\mu'} &=& \sum\limits_{12}U^{\dagger}_{\mu 1}f_{12}U_{2\mu'} \ \ \ \ \ \ \  {\tilde F}^{11}_{\mu\mu'} = -\sum\limits_{12}V^{T}_{\mu 1}f_{12}V^{\ast}_{2\mu'}  \nonumber\\
F^{00} &=& \sum\limits_{12}V^{T}_{\mu 1}f_{12}V^{\ast}_{2\mu}.
 \eea

The  $F^{11}$ and ${\tilde F}^{11}$ terms are usually dropped in the zero-temperature QRPA formalism \cite{Avogadro2011}, however, they play a non-negligible role at finite temperatures and in QRPA extensions, therefore, we keep them here to maintain the full component structure of the matrix elements. The $F^{00}$ term does non induce transitions, so it can be dropped henceforth.
%the fermionic nature of the operators allows one to use reduced summations:
%\bea
%F = \sum\limits_{\mu\leq\mu'} \Bigl(F^{20}_{\mu\mu'}\alpha^{\dagger}_{\mu}\alpha^{\dagger}_{\mu'} + 
%F^{02}_{\mu\mu'}\alpha_{\mu'}\alpha_{\mu} + F^{11}_{\mu\mu'}\alpha^{\dagger}_{\mu}\alpha_{\mu'}  \nonumber \\
%+ {\tilde F}^{11}_{\mu\mu'}\alpha^{\dagger}_{\mu'}\alpha_{\mu} \Bigr) \ \ \ \ \ \ \ 
%\nonumber\\
%\label{Fextred}
%\eea
The transition probabilities in the quasiparticle picture read:
\bea
|\langle n|F^{\dagger}|0\rangle |^2 = \frac{1}{4}\sum\limits_{\mu\mu'\nu\nu'}
\left(\begin{array}{cccc} F^{02}_{\mu\mu'} & F^{20}_{\mu\mu'} & F^{11}_{\mu\mu'}& {\tilde F}^{11}_{\mu\mu'}\end{array}\right) \nonumber\\
\times\left(\begin{array}{cccc} X^{n}_{\mu\mu'}X^{n\ast}_{\nu\nu'}  &  X^{n}_{\mu\mu'}Y^{n\ast}_{\nu\nu'}
&  X^{n}_{\mu\mu'}U^{n\ast}_{\nu\nu'}&  X^{n}_{\mu\mu'}V^{n\ast}_{\nu\nu'}
\\[3pt]
Y^{n}_{\mu\mu'}X^{n\ast}_{\nu\nu'}  &  Y^{n}_{\mu\mu'}Y^{n\ast}_{\nu\nu'}
&Y^{n}_{\mu\mu'}U^{n\ast}_{\nu\nu'}  &  Y^{n}_{\mu\mu'}V^{n\ast}_{\nu\nu'}
\\[3pt]
U^{n}_{\mu\mu'}X^{n\ast}_{\nu\nu'}  &  U^{n}_{\mu\mu'}Y^{n\ast}_{\nu\nu'}
&U^{n}_{\mu\mu'}U^{n\ast}_{\nu\nu'}  &  U^{n}_{\mu\mu'}V^{n\ast}_{\nu\nu'}
\\[3pt]
V^{n}_{\mu\mu'}X^{n\ast}_{\nu\nu'}  &  V^{n}_{\mu\mu'}Y^{n\ast}_{\nu\nu'}
&V^{n}_{\mu\mu'}U^{n\ast}_{\nu\nu'}  &  V^{n}_{\mu\mu'}V^{n\ast}_{\nu\nu'}
\end{array}\right)
\left(\begin{array}{c} F^{02\ast}_{\nu\nu'} \\[3pt] F^{20\ast}_{\nu\nu'} \\[3pt] F^{11\ast}_{\nu\nu'} \\[3pt] {\tilde F}^{11\ast}_{\nu\nu'}\end{array}\right)\nonumber\\
\label{nFd0}
\eea
via the matrix elements
\bea
X^{n}_{\mu\mu'} = \langle 0|\alpha_{\mu'}\alpha_{\mu}|n\rangle \ \ \ \ \ Y^{n}_{\mu\mu'} = \langle 0|\alpha^{\dagger}_{\mu}\alpha^{\dagger}_{\mu'}|n\rangle \nonumber\\
U^{n}_{\mu\mu'} = \langle 0|\alpha^{\dagger}_{\mu}\alpha_{\mu'}|n\rangle \ \ \ \ \ V^{n}_{\mu\mu'} = \langle 0|\alpha^{\dagger}_{\mu'}\alpha_{\mu}|n\rangle .\nonumber \\
\label{XY}
\eea
Further, we assign special notations to the quasiparticle pair operators that enter the operator expansion (\ref{Fext}), in agreement with Ref. \cite{Litvinova2022},
\bea
A_{\mu\mu'} = \alpha_{\mu'}\alpha_{\mu} \ \ \ \ \ \ \ \ \ \ A^{\dagger}_{\mu\mu'} = \alpha^{\dagger}_{\mu}\alpha^{\dagger}_{\mu'}\nonumber\\
C_{\mu\mu'} = \alpha^{\dagger}_{\mu}\alpha_{\mu'} \ \ \ \ \ \ \ \ \ \ C^{\dagger}_{\mu\mu'} = \alpha^{\dagger}_{\mu'}\alpha_{\mu},
\label{Aq}
\eea
unified notations for the external field operator components
\be
F^{02} = {\cal F}^{[1]},\ \ \ \ \ F^{20} = {\cal F}^{[2]},\ \ \ \ \ F^{11} = {\cal F}^{[3]},\ \ \ \ \ {\tilde F}^{11} = {\cal F}^{[4]}
\ee 
and matrix elements
\bea
Z^{n[1+]}_{\mu\mu'} &=& X^{n}_{\mu\mu'}  \ \ \ \ \ Z^{n[2+]}_{\mu\mu'} = Y^{n}_{\mu\mu'} \nonumber\\
Z^{n[3+]}_{\mu\mu'} &=& U^{n}_{\mu\mu'}  \ \ \ \ \ Z^{n[4+]}_{\mu\mu'} = V^{n}_{\mu\mu'} \nonumber \\
Z^{n[1-]}_{\mu\mu'} &=& Y^{n\ast}_{\mu\mu'}  \ \ \ \ \ Z^{n[2-]}_{\mu\mu'} = X^{n\ast}_{\mu\mu'} \nonumber\\
Z^{n[3-]}_{\mu\mu'} &=& V^{n\ast}_{\mu\mu'}  \ \ \ \ \ Z^{n[4-]}_{\mu\mu'} = U^{n\ast}_{\mu\mu'}. \nonumber \\
\label{XYZ}
\eea

The $\delta$-functions in Eq. (\ref{SF}) can be represented as zero-width limits of the Lorentzian distribution
\be
\delta(\omega-\omega_n) = \frac{1}{\pi}\lim\limits_{\Delta \to 0}\frac{\Delta}{(\omega - \omega_n)^2 + \Delta^2},
\ee
so that
\bea
S(\omega) = -\frac{1}{\pi}\lim\limits_{\Delta \to 0} \text{Im} \Pi(\omega+i\Delta),
\label{SFDelta} 
\eea
with the polarizability $\Pi(\omega)$ of the system defined as
\bea
\Pi(\omega+i\Delta) = \sum\limits_{n>0} \Bigl[ \frac{|\langle n|F^{\dagger}|0\rangle |^2}{\omega - \omega_n + i\Delta}
- \frac{|\langle n|F|0\rangle |^2}{\omega + \omega_n + i\Delta}
\Bigr] \nonumber\\
 = \frac{1}{4}\sum\limits_{i,j=1}^{4}\sum\limits_{\mu\mu'\nu\nu'}{\cal F}^{[i]}_{\mu\mu'} R^{[ij]}_{\mu\mu'\nu\nu'}(\omega+i\Delta) 
{\cal F}^{[j]\ast}_{\nu\nu'},\nonumber\\
\label{Polar}
\eea
where the matrix elements of the $4\times 4$ block matrix response function $R^{[ij]}_{\mu\mu'\nu\nu'}(\omega)$ read:
\bea
R^{[ij]}_{\mu\mu'\nu\nu'}(\omega) &=& \sum\limits_{n>0} \Bigl(
\frac{Z^{n[i+]}_{\mu\mu'} Z^{n[j+]\ast}_{\nu\nu'}}{\omega - \omega_n} - 
\frac{Z^{n[j-]}_{\mu\mu'} Z^{n[i-]\ast}_{\nu\nu'}}{\omega + \omega_n}\Bigr)\nonumber\\
&=& \sum\limits_{n>0}\sum\limits_{\sigma=\pm} \sigma
\frac{Z^{n[i\sigma]}_{\mu\mu'} Z^{n[j\sigma]\ast}_{\nu\nu'}}{\omega - \sigma\omega_n} .
\label{Romega}
\eea
Similarly to the logic of Ref. \cite{Litvinova2022}, we want to relate the response function to a two-time two-body propagator, but now in the extended $4\times 4$ form. 
The definition compatible with Eq. (\ref{Romega}), should be its Fourier image. Thus, we adopt the form
\be
{\hat R}_{\mu\mu'\nu\nu'} (t-t') = -i\langle T{\hat\Psi}_{\mu\mu'}(t){\hat\Psi}^{\dagger}_{\nu\nu'}(t')\rangle
\label{Rt}
\ee
with the four-component vector 
\be
{\hat\Psi}_{\mu\mu'} (t) = \left(\begin{array}{c}
A_{\mu\mu'}(t) \\[3pt] A^{\dagger}_{\mu\mu'}(t) \\[3pt] C_{\mu\mu'}(t) \\[3pt] C^{\dagger}_{\mu\mu'}(t)
\end{array}\right) \equiv \left(\begin{array}{c}
\Psi^{[1]}_{\mu\mu'}(t) \\[3pt] \Psi^{[2]}_{\mu\mu'}(t) \\[3pt] \Psi^{[3]}_{\mu\mu'}(t) \\[3pt] \Psi^{[4]}_{\mu\mu'}(t)
\end{array}\right)
\label{Psi}
\ee
and the time-dependent pair operators in the Heisenberg picture:
\bea
A_{\mu\mu'}(t) = e^{iHt}\alpha_{\mu'}\alpha_{\mu}e^{-iHt} \nonumber \\ 
A^{\dagger}_{\mu\mu'}(t) = e^{iHt}\alpha^{\dagger}_{\mu}\alpha^{\dagger}_{\mu'}e^{-iHt}\nonumber \\ 
C_{\mu\mu'}(t) = e^{iHt}\alpha^{\dagger}_{\mu}\alpha_{\mu'}e^{-iHt} \nonumber \\ 
C^{\dagger}_{\mu\mu'}(t) = e^{iHt}\alpha^{\dagger}_{\mu'}\alpha_{\mu}e^{-iHt},
\eea 
with the $\hbar = 1$ convention, $T$ is the time ordering operator, and the averaging is performed over the formally exact ground state. Inserting the operator unit $I = \sum_n|n\rangle\langle n|$, in terms of the full set of the many-body Hamiltonian eigenstates, between the $t$ and $t'$-dependent operators in Eq. (\ref{Rt}) and performing the Fourier transformation to the frequency (energy) domain, we arrive at:
\bea
R^{[ij]}_{\mu\mu'\nu\nu'}(\omega) = \sum\limits_{n>0}\sum\limits_{\sigma=\pm} \sigma
\frac{Z^{n[i\sigma]}_{\mu\mu'} Z^{n[j\sigma]\ast}_{\nu\nu'}}{\omega - \sigma(\omega_n - i\delta)}, 
\label{Romega1}
\eea
where $\delta\to +0$ is the infinitesimal imaginary part of the energy variable introduced for convergence of the integrals over $\tau = t-t'$ for the retarded and advanced contributions separately. Thus, the definition (\ref{Rt}) is consistent with Eq. (\ref{Romega}), if the different origins of the imaginary parts $\Delta$ and $\delta$ of the energy variable are recognized.
Once the response function is linked to the two-quasiparticle ($2q$) propagator (\ref{Rt}), the application of the EOM method is straightforward.

%\subsection{Equation of motion for the superfluid response}

Thanks to the block matrix notations, the EOM for ${\hat R}$ can be derived keeping the $4\times 4$ structure consistently.
Namely, we consider the derivative with respect to the first time argument:
\bea
\partial_t{\hat R}_{\mu\mu'\nu\nu'}(t-t') = -i\delta(t-t')
\langle\left[{\hat\Psi}_{\mu\mu'},{\hat\Psi}^{\dagger}_{\nu\nu'}\right]\rangle \nonumber \\ 
+ \langle T\left[H,{\hat\Psi}_{\mu\mu'}\right](t){\hat\Psi}^{\dagger}_{\nu\nu'}(t')\rangle ,
\label{dR0}
\eea
where we adopted the following notation for operator products and commutators:
\be
[{H},A](t) = e^{i{H}t}[{H},A]e^{-i{H}t}
\ee
for an arbitrary single-time operator (product) $A$. 
%so that one can immediately recognize the 
At this step, we evaluate explicitly the commutators with the one-quasiparticle part of the full Hamiltonian (\ref{Hqua1}):
\bea
[H,A_{\mu\mu'}] &=& -(E_{\mu} + E_{\mu'})A_{\mu\mu'} + [V,A_{\mu\mu'}]\nonumber\\
\left[H,A^{\dagger}_{\mu\mu'}\right] &=& (E_{\mu} + E_{\mu'} )A^{\dagger}_{\mu\mu'} + [V,A^{\dagger}_{\mu\mu'}]\nonumber\\
\left[H,C_{\mu\mu'}\right] &=& (E_{\mu} - E_{\mu'})C_{\mu\mu'} + [V,C_{\mu\mu'}]\nonumber\\
\left[H,C^{\dagger}_{\mu\mu'}\right] &=& -(E_{\mu} - E_{\mu'} )C^{\dagger}_{\mu\mu'} + [V,C^{\dagger}_{\mu\mu'}],
\label{Comm1}
\eea
or, in the matrix form,
\be
[H,{\hat\Psi}_{{\mu\mu'}}] = -{\hat\Sigma}_3{\hat E}_{\mu\mu'} + [V,{\hat\Psi}_{{\mu\mu'}}]
\ee
with the $4\times 4$ block matrices
\bea
{\hat\Sigma}_3 &=& \left(\begin{array}{cc} \sigma_3& 0 \\ 0 & -\sigma_3 \end{array}\right) \ \ \ \ \ \ \ \ \ \ 
\sigma_3 = \left(\begin{array}{cc}1& 0 \\ 0 & -1 \end{array}\right)
 \nonumber \\ 
{\hat E}_{\mu\mu'} &=& \left(\begin{array}{cc} (E_{\mu} + E_{\mu'})\times \mathbb{1} & 0 \\ 0 & (E_{\mu} - E_{\mu'})\times \mathbb{1} \end{array}\right).
\eea
After that, Eq. (\ref{dR0}) leads to the first EOM:
\bea
(i\partial_t - {\hat\Sigma}_3{\hat E}_{\mu\mu'}){\hat R}_{\mu\mu'\nu\nu'}(t-t') 
= \delta(t-t'){\hat N}_{\mu\mu'\nu\nu'}  \nonumber\\
+ i \langle T[V,{\hat\Psi}_{\mu\mu'}](t){\hat\Psi}^{\dagger}_{\nu\nu'}(t')\rangle , \ \ \ \ \ \ \ \ \ \ 
\label{EOM1}
\eea
where the norm matrix ${\hat N}_{\mu\mu'\nu\nu'}$ is compactly defined as:
\be
{\hat N}_{\mu\mu'\nu\nu'} = \langle[{\hat\Psi}_{\mu\mu'},{\hat\Psi}^{\dagger}_{\nu\nu'}]\rangle .
\label{norm}
\ee
In analogy to the normal case, the second EOM addresses the time-dependent term on the right hand side of Eq. (\ref{EOM1}):
\bea
{\hat{F}}_{\mu\mu'\nu\nu'}(t-t') =
i \langle T[V,{\hat\Psi}_{\mu\mu'}](t){\hat\Psi}^{\dagger}_{\nu\nu'}(t')\rangle,
\label{F}
\eea
and its derivative is taken now with respect to $t'$:
\bea
\partial_{t'}{\hat{F}}_{\mu\mu'\nu\nu'}(t-t') &=& -i\delta(t-t')
\langle\left[[V,{\hat\Psi}_{\mu\mu'}],{\hat\Psi}^{\dagger}_{\nu\nu'}\right]\rangle \nonumber\\
&-& \langle T[V,{\hat\Psi}_{\mu\mu'}](t)[H,{\hat\Psi}^{\dagger}_{\nu\nu'}](t')  \rangle . 
\label{dR}
\eea
Processing the commutators with the Hamiltonian as in Eq. (\ref{Comm1}) brings the second EOM to the form:
\bea
{\hat F}_{\mu\mu'\nu\nu'}(t&-&t') [-i\overleftarrow{\partial_{t'}} - {\hat\Sigma}_3{\hat E}_{\nu\nu'}]
=
 \nonumber\\
&=& \delta(t-t'){\hat T}^{0}_{\mu\mu'\nu\nu'} + {\hat T}^{r}_{\mu\mu'\nu\nu'}(t-t'),
\label{EOM2}
\eea
where ${\hat T}^{0}_{\mu\mu'\nu\nu'}$  and ${\hat T}^{r}_{\mu\mu'\nu\nu'}(t-t')$ introduce the generalized static and dynamical (retarded, or time-dependent) parts of the 
two-fermion ${\hat T}$-matrix in the quasiparticle space:
\bea
{\hat T}^{0}_{\mu\mu'\nu\nu'} &=& -\langle\left[[V,{\hat\Psi}_{\mu\mu'}],{\hat\Psi}^{\dagger}_{\nu\nu'}\right]  \rangle \\
\label{T0}
{\hat T}^{r}_{\mu\mu'\nu\nu'}(t-t') &=& i\langle T[V,{\hat\Psi}_{\mu\mu'}](t)[V,{\hat\Psi}^{\dagger}_{\nu\nu'}](t')\rangle . 
\label{Tr}
\eea
Acting by the operator, standing in the square brackets in the left-hand side of Eq. (\ref{EOM2}), on Eq. (\ref{EOM1}) leads to the EOM for the quasiparticle propagator:
\bea
\left[i\partial_t \right. &-& \left.{\hat\Sigma}_3{\hat E}_{\mu\mu'}\right]{\hat R}_{\mu\mu'\nu\nu'}(t-t') \left[-i\overleftarrow{\partial_{t'}} - {\hat\Sigma}_3{\hat E}_{\nu\nu'}\right] \nonumber\\
&=& \delta(t-t'){\hat N}_{\mu\mu'\nu\nu'}\left[-i\overleftarrow{\partial_{t'}} - {\hat\Sigma}_3{\hat E}_{\nu\nu'}\right] \nonumber\\
&+&  \delta(t-t'){\hat T}^{0}_{\mu\mu'\nu\nu'} + {\hat T}^{r}_{\mu\mu'\nu\nu'}(t-t'),
\label{EOMcomb}
\eea
whose Fourier transformation yields
\bea
\left[\omega\right. &-& \left.{\hat\Sigma}_3{\hat E}_{\mu\mu'}\right]{\hat R}_{\mu\mu'\nu\nu'}(\omega) \left[\omega\right. - \left.{\hat\Sigma}_3{\hat E}_{\nu\nu'}\right] \nonumber\\
&=& {\hat N}_{\mu\mu'\nu\nu'}\left[\omega - {\hat\Sigma}_3{\hat E}_{\nu\nu'}\right] 
+  {\hat T}^{0}_{\mu\mu'\nu\nu'} + {\hat T}^{r}_{\mu\mu'\nu\nu'}(\omega). \nonumber\\
\label{EOMcomb_omega}
\eea
The generalized uncorrelated two-quasiparticle propagator is naturally defined as
\be
{\hat R}^0_{\mu\mu'\nu\nu'}(\omega) = \left[\omega\right. - \left.{\hat\Sigma}_3{\hat E}_{\mu\mu'}\right]^{-1}{\hat N}_{\mu\mu'\nu\nu'},
\ee
so that Eq. (\ref{EOMcomb_omega}) becomes a recognizable $T$-matrix equation:
\bea
{\hat R}_{\mu\mu'\nu\nu'}(\omega) = {\hat R}^0_{\mu\mu'\nu\nu'}(\omega)  \ \ \ \ \ \ \ \ \ \ \ \ \ \nonumber\\
+ 
\frac{1}{4}\sum\limits_{\gamma\gamma'\delta\delta'}
{\hat R}^0_{\mu\mu'\gamma\gamma'}(\omega){\hat T}_{\gamma\gamma'\delta\delta'}(\omega){\hat R}^0_{\delta\delta'\nu\nu'}(\omega) \ \ \ \ \ 
\eea
with the energy-dependent $T$-matrix ${\hat T}_{\gamma\gamma'\delta\delta'}(\omega)$ such, that
\be
{\hat T}_{\gamma\gamma'\delta\delta'}(\omega) = \frac{1}{4}\sum\limits_{\mu\mu'\nu\nu'}{\hat N}^{-1}_{\gamma\gamma'\mu\mu'}
\bigl({\hat T}^{0}_{\mu\mu'\nu\nu'} + {\hat T}^{r}_{\mu\mu'\nu\nu'}(\omega)\bigr)
{\hat N}^{-1}_{\nu\nu'\delta\delta'},
\ee
while we assume that the inverse norm matrix satisfies:
\bea
\frac{1}{2}\sum_{\delta\delta'}{\hat N}^{-1}_{\mu\mu'\delta\delta'}{\hat N}_{\delta\delta'\nu\nu'} = \delta_{\mu\mu'\nu\nu'} =
 \delta_{\mu\nu}\delta_{\mu'\nu'} - \delta_{\mu\nu'}\delta_{\mu'\nu}.\nonumber\\
\eea
Transformation of the $T$-matrix equation to the Dyson-Bethe-Salpeter equation (Dyson-BSE) form can be performed by introducing the interaction kernel ${\hat K}(\omega) = {\hat K}^0 + {\hat K}^r(\omega)$, irreducible with respect to ${\hat R}^0$, i.e. ${\hat K}$ does not contain terms partitioned by the uncorrelated propagator ${\hat R}^0$.
The interaction kernel plays the role of the self-energy for the two-point two-quasiparticle correlation function:
\bea
{\hat K}^0_{\gamma\gamma'\delta\delta'} = \frac{1}{4}\sum\limits_{\eta\eta'\rho\rho'}{\hat N}^{-1}_{\gamma\gamma'\eta\eta'}
{\hat T}^{0}_{\eta\eta'\rho\rho'} 
{\hat N}^{-1}_{\rho\rho'\delta\delta'} \nonumber\\
{\hat K}^r_{\gamma\gamma'\delta\delta'}(\omega) = \frac{1}{4}\sum\limits_{\eta\eta'\rho\rho'}\left[{\hat N}^{-1}_{\gamma\gamma'\eta\eta'}
{\hat T}^{r}_{\eta\eta'\rho\rho'}(\omega) 
{\hat N}^{-1}_{\rho\rho'\delta\delta'}\right]^{irr}. \nonumber\\
\eea
Finally, Dyson-BSE can be packed into the familiar form
\bea
{\hat R}_{\mu\mu'\nu\nu'}(\omega) = {\hat R}^0_{\mu\mu'\nu\nu'}(\omega) \nonumber\\
+ \frac{1}{4}\sum\limits_{\gamma\gamma'\delta\delta'}{\hat R}^0_{\mu\mu'\gamma\gamma'}(\omega){\hat K}_{\gamma\gamma'\delta\delta'}(\omega){\hat R}_{\delta\delta'\nu\nu'}(\omega),
\label{BSDE}
\eea
but with the $4\times 4$ matrix structure in the quasiparticle basis.

\subsubsection{Finite temperature}

The logic of defining the response function at finite temperature is similar but slightly different from that at zero temperature. The starting point is dictated by the modified definition of the strength function
\cite{RingRobledoEgidoEtAl1984}
\begin{eqnarray}
\tilde{S}(\omega)&=&\sum_{if}p_{i}|\langle f|F^{\dag}|i\rangle|^{2}\delta(\omega-E_{f}+E_{i})\equiv S_{+}(\omega), \nonumber\\
\end{eqnarray}
where $f$ denotes the set of final states and $i$ run over all possible initial states distributed with the probabilities $p_{i}$ of finding the system in those states within the grand canonical ensemble \cite{Goodman1981,Sommermann1983}:
\begin{equation}
p_{i}=\frac{e^{-E_{i}/T}}{\sum_{j}e^{-E_{j}/T}},
\end{equation}
and $T$ stands for the temperature.
In this case, there is no formal borderline between absorption and emission because the transition frequency $\omega_{fi} = E_f - E_i$ can be both positive and negative.  
Using the principle of detailed balance, the opposite frequency strength function can be defined: 
\begin{equation}
S_{-}(\omega)=\sum_{if}p_{i}|\langle f|F|i\rangle|^{2}\delta(\omega+E_{f}-E_{i})
\end{equation}
and related to the $S_{+}$ as $S_{-}(\omega) = e^{-\omega/T}S_{+}(\omega)$ by interchanging $i\leftrightarrow f$ under summation,
so that 
\begin{eqnarray}
{\tilde S}(\omega)&=&\frac{1}{1-e^{-\omega/T}}\left[S_{+}(\omega)-S_{-}(\omega)\right]\nonumber\\
&=&-\frac{1}{\pi}\frac{1}{1-e^{-\omega/T}}\lim_{\Delta\rightarrow+0}\text{Im}\sum_{if}p_{i}
%\left[
\Bigl(\frac{|\langle f|F^{\dagger}|i\rangle|^{2}}{\omega-E_{f}+E_{i}+i\Delta}\nonumber\\
&-& \frac{|\langle f|F|i\rangle|^{2}}{\omega+E_{f}-E_{i}+i\Delta}\Bigr).
%\right] . %\nonumber\\
\end{eqnarray}
This allows for identifying the finite-temperature response function in analogy to the $T=0$ case:
\bea
{\cal R}^{[lk]}_{\mu\mu'\nu\nu'}(\omega) = \sum\limits_{fi}\sum\limits_{\sigma=\pm} \sigma p_i
\frac{{\cal Z}^{if[l\sigma]}_{\mu\mu'} {\cal Z}^{if[k\sigma]\ast}_{\nu\nu'}}{\omega - \sigma\omega_{fi}}, 
\label{Romega1T}
\eea
where
\bea
{\cal Z}^{if[1+]}_{\mu\mu'} &=& {\cal X}^{if}_{\mu\mu'}  \ \ \ \ \ {\cal Z}^{if[2+]}_{\mu\mu'} = {\cal Y}^{if}_{\mu\mu'} \nonumber\\
{\cal Z}^{if[3+]}_{\mu\mu'} &=& {\cal U}^{if}_{\mu\mu'}  \ \ \ \ \ {\cal Z}^{if[4+]}_{\mu\mu'} = {\cal V}^{if}_{\mu\mu'} \nonumber \\
{\cal Z}^{if[1-]}_{\mu\mu'} &=& {\cal Y}^{if\ast}_{\mu\mu'}  \ \ \ \ \ {\cal Z}^{if[2-]}_{\mu\mu'} = {\cal X}^{if\ast}_{\mu\mu'} \nonumber\\
{\cal Z}^{if[3-]}_{\mu\mu'} &=& {\cal V}^{if\ast}_{\mu\mu'}  \ \ \ \ \ {\cal Z}^{if[4-]}_{\mu\mu'} = {\cal U}^{if\ast}_{\mu\mu'} \nonumber \\
\label{XYZt}
\eea
and 
\bea
{\cal X}^{if}_{\mu\mu'} = \langle i|\alpha_{\mu'}\alpha_{\mu}|f\rangle \ \ \ \ \ {\cal Y}^{if}_{\mu\mu'} = \langle i|\alpha^{\dagger}_{\mu}\alpha^{\dagger}_{\mu'}|f\rangle \nonumber\\
{\cal U}^{if}_{\mu\mu'} = \langle i|\alpha^{\dagger}_{\mu}\alpha_{\mu'}|f\rangle \ \ \ \ \ {\cal V}^{if}_{\mu\mu'} = \langle i|\alpha^{\dagger}_{\mu'}\alpha_{\mu}|f\rangle .\nonumber \\
\label{XYUV}
\eea
The spectral part of the strength function can be defined as
\bea
S(\omega) = -\frac{1}{\pi}\lim\limits_{\Delta \to 0} \text{Im} {\Pi}(\omega+i\Delta),
\label{SFDeltaT} 
\eea
with the polarizability $\Pi(\omega)$, in analogy to the zero-temperature case, 
\bea
\Pi(\omega+i\Delta)
 = \frac{1}{4}\sum\limits_{i,j=1}^{4}\sum\limits_{\mu\mu'\nu\nu'}{\cal F}^{[i]}_{\mu\mu'} {\cal R}^{[ij]}_{\mu\mu'\nu\nu'}(\omega+i\Delta) 
{\cal F}^{[j]\ast}_{\nu\nu'}.\nonumber\\
\label{PolarT}
\eea
%%%%%%% Check Eq. above, introduce and differentiate FT notations <f|...|i>
%In terms of the finite-temperature response function ($\delta\rightarrow+0$):
%\begin{eqnarray}
%\mathcal{R}_{k_{1}k_{2},k_{3}k_{4}}(E)&=&\sum_{if}p_{i}
%\left\{
%\frac{\langle f|\hat{a}^{\dag}_{k_{4}}\hat{a}_{k_{3}}|i\rangle\langle i|\hat{a}^{\dag}_{k_{1}}\hat{a}_{k_{2}}|f\rangle}{E+E_{f}-E_{i}+i\delta}-\nonumber\\
%&-&\frac{\langle f|\hat{a}^{\dag}_{k_{1}}\hat{a}_{k_{2}}|i\rangle\langle i|\hat{a}^{\dag}_{k_{4}}\hat{a}_{k_{3}}|f\rangle}{E-E_{f}+E_{i}+i\delta}%\right\},
%\end{eqnarray}
The strength function $\tilde{S}(E)$ can be then expressed as
\begin{eqnarray}
\label{Strength}
\tilde{S}(E)&=&\frac{1}{1-e^{-E/T}}{S}(E),
\end{eqnarray}
i.e., has an additional factor $\left[1-\exp(-E/T)\right]^{-1}$, singular at $E=0$, as compared to the zero-temperature strength function.  
%is, thereby, the new feature which is inherent for the finite-temperature strength function. In particular, it influences the low-energy behavior of $\tilde{S}(E)$, in addition to the appearance %of new poles in the response function, and makes the zero-energy limit of $\tilde{S}(E)$ finite at $T>0$, in contrast to that of the spectral density  ${S}(E)$, whose zero-energy limit is zero %at all temperatures.
%%%%

The field-theoretical part of the temperature-dependent formalism can be conveniently carried out following Matsubara \cite{Matsubara1955}. The propagator of a fermionic quasiparticle pair in a heated correlated medium is the thermal analog of Eq. (\ref{Rt}) defined with thermal average \cite{Zagoskin2014}:
\be
{\hat {\cal R}}_{\mu\mu'\nu\nu'}(\tau-\tau') = -\langle {\cal T} {\hat\Psi}_{\mu\mu'}(\tau){\hat{\bar\Psi}}_{\nu\nu'}(\tau')\rangle 
\label{RtT} 
\ee
employing the chronological ordering $\cal T$ of the quasiparticle four-component field operators in the imaginary time domain of the Wick rotated picture:
\bea
{\hat\Psi}_{\mu\mu'}(\tau) &=& e^{{\cal H}\tau}{\hat\Psi}_{\mu\mu'}e^{-{\cal H}\tau}\nonumber\\
{\hat{\bar\Psi}}_{\mu\mu'}(\tau) &=& e^{{\cal H}\tau}{\hat{\Psi}^{\dagger}}_{\mu\mu'}e^{-{\cal H}\tau}.
\label{Wick-Heisenberg}
\eea
The operator ${\cal H}$ is the redefined Hamiltonian ${\cal H} = H - \lambda N$, where $\lambda$ is the chemical potential and $N$ is the particle number operator, while the angular brackets in Eq. (\ref{RtT}) stand for the thermal average \cite{Abrikosov1975, Zagoskin2014}
\be
\langle O \rangle = \sum\limits_{i} p_{i}\langle i|O|i\rangle .
\ee

% Here comes the Fourier transform and comparison with the above SF

The time evolution of the correlated fermionic quasiparticle pair, or propagator, defined by Eq. (\ref{RtT}) with the imaginary times, is generated by the differentiation of this function with respect to the first time variable $\tau$:
\bea
\partial_{\tau} {\hat{\cal R}}_{\mu\mu'\nu\nu'}(\tau-\tau') = -\delta(\tau-\tau')\langle [{\hat\Psi}_{\mu\mu},{\hat{\bar\Psi}}_{\nu\nu'}]\rangle -\nonumber \\
- \langle {\cal T}[{\cal H},{\hat\Psi}_{\mu\mu}](\tau)({\hat{\bar\Psi}}_{\nu\nu}(\tau')\rangle, \nonumber\\ 
\label{dtR}                           
\eea
where, analogously to the previous paragraph,
\be
[{\cal H},A](\tau) = e^{{\cal H}\tau}[{\cal H},A]e^{-{\cal H}\tau}.
\ee
After evaluating the commutators with the one-body parts of the Hamiltonian,
the first EOM reads:
\bea
(\partial_{\tau} + {\hat\Sigma}_3{\hat E}_{\mu\mu'}){\hat {\cal R}}_{\mu\mu'\nu\nu'}(\tau-\tau') 
= -\delta(\tau-\tau'){\hat {\cal N}}_{\mu\mu'\nu\nu'}  \nonumber\\
- \langle {\cal T}[V,{\hat\Psi}_{\mu\mu'}](\tau){\hat{\bar\Psi}}_{\nu\nu'}(\tau')\rangle , \ \ \ \ \ \ \ \ \ \ 
\label{EOM1T}
\eea
with the norm matrix
\be
{\hat {\cal N}}_{\mu\mu'\nu\nu'} = \langle [{\hat\Psi}_{\mu\mu},{\hat{\bar\Psi}}_{\nu\nu'}]\rangle ,
\label{normT}
\ee
which is the finite-temperature cousin of the norm of Eq. (\ref{norm}). Again, we isolate the second term on the right-hand side of Eq. (\ref{EOM1T})
\be
{\hat{\cal F}}_{\mu\mu'\nu\nu'}(\tau-\tau') = \langle {\cal T}[V,{\hat\Psi}_{\mu\mu'}](\tau){\hat{\bar\Psi}}_{\nu\nu'}(\tau')\rangle 
\ee
for the second EOM
 with respect to the second time argument $\tau'$:
\bea
{\hat{\cal F}}_{\mu\mu'\nu\nu'}(\tau-\tau')(\overleftarrow{\partial}_{\tau'} &-& {\hat\Sigma}_3{\hat E}_{\nu\nu'}) \nonumber\\ 
= -\delta(\tau-\tau')\langle[[V,{\hat\Psi}_{\mu\mu}],{\hat{\bar\Psi}}_{\nu\nu'}]\rangle %+ \nonumber\\ 
&+& \langle  {\cal T}[V,{\hat\Psi}_{\mu\mu}](\tau)[V,{\hat{\bar\Psi}}_{\nu\nu'}](\tau')\rangle .\nonumber\\
\label{EOM2T}
\eea
Applying the operator $(\overleftarrow{\partial}_{\tau'} - {\hat\Sigma}_3{\hat E}_{\nu\nu'})$ to the first EOM (\ref{EOM1T}) allows combining Eqs. (\ref{EOM1T}) and (\ref{EOM2T}). The Fourier transformation to the domain of discrete energy variable $\omega_n = 2\pi i nT$ is defined as:
% i removed from the exp, included in \omega_n instead
\be
{\hat{\cal R}}_{\mu\mu'\nu\nu'}(\tau-\tau') = T\sum\limits_{n}  e^{-\omega_n(\tau-\tau')}{\hat{\cal R}}_{\mu\mu'\nu\nu'}(\omega_n).
\ee 
In this way, we obtain:
\bea
\hat{\cal R}_{\mu\mu'\nu\nu'}(\omega_n) &=& \hat{\cal R}^{0}_{\mu\mu'\nu\nu'}(\omega_n)  \nonumber \\
+&\  &\frac{1}{4}\sum\limits_{\gamma\gamma'\delta\delta'}\hat{\cal R}^{0}_{\mu\mu'\gamma\gamma'}(\omega_n)
{\hat{\cal T}}_{\gamma\gamma'\delta\delta'}(\omega_n)\hat{\cal R}^{0}_{\delta\delta'\nu\nu'}(\omega_n),\nonumber\\  
\label{RTmatrixT}
\eea
where the free propagator is introduced as:
\be
\hat{\cal R}^{0}_{\mu\mu'\nu\nu'}(\omega_n ) = [\omega_n -  {\hat\Sigma}_3{\hat E}_{\mu\mu'}]^{-1}{\hat{\cal N}}_{\mu\mu'\nu\nu'}
\label{R0}
\ee
and the interaction, or $T$-matrix, reads:
\be
{\hat {\cal T}}_{\mu\mu'\nu\nu'}(\omega_n) =  \frac{1}{4}\sum\limits_{\gamma\gamma'\delta\delta'}{\hat {\cal N}}^{-1}_{\mu\mu'\gamma\gamma'}\Bigl( {\hat{\cal T}}^{0}_{\gamma\gamma'\delta\delta'} + {\hat{\cal T}}^{r}_{\gamma\gamma'\delta\delta'}(\omega_n)\Bigr){\hat{\cal N}}^{-1}_{\delta\delta'\nu\nu'},
\label{TN}
\ee
being the Fourier transform of
\bea
{\hat{\cal T}}^{0}_{\mu\mu'\nu\nu'}(\tau-\tau') &=&  -\delta(\tau-\tau') \langle  [[V,{\hat\Psi}_{\mu\mu'}],{\hat{\bar\Psi}}_{\nu\nu'}]\rangle\nonumber\\
{\hat{\cal T}}^{r}_{\mu\mu'\nu\nu'}(\tau-\tau') &=&  \langle  {\cal T}[V,{\hat\Psi}_{\mu\mu'}](\tau)[V,{\hat{\bar\Psi}}_{\nu\nu'}](\tau')\rangle . \nonumber\\
\label{Tt}
\eea
Thus, the $T$-matrix maintains its split into the instantaneous, or static, part ${\hat{\cal T}}^{0}$ and time-dependent, or dynamic(al), component ${\hat{\cal T}}^{r}$ also at finite temperature and in the presence of superfluidity.
The Dyson-BSE corresponding to Eq. (\ref{RTmatrixT}) reads:
\bea
{\hat{\cal R}}_{\mu\mu'\nu\nu'}(\omega_n) &=& {\hat{\cal R}}^{0}_{\mu\mu'\nu\nu'}(\omega_n) + \nonumber \\
&+& \frac{1}{4}\sum\limits_{\gamma\gamma'\delta\delta'}{\hat{\cal R}}^{0}_{\mu\mu'\gamma\gamma'}(\omega_n){\hat{\cal K}}_{\gamma\gamma'\delta\delta'}(\omega_n){\hat{\cal R}}_{\delta\delta'\nu\nu'}(\omega_n)\nonumber\\  
\label{RDyson}
\eea
with the kernel ${\hat{\cal K}}$ irreducible with respect to ${\hat{\cal R}}^{0}$, or two-quasiparticle analog of the self-energy:
\bea
{\hat{\cal T}}_{\mu\mu'\nu\nu'}(\omega_n) &=& {\hat{\cal K}}_{\mu\mu'\nu\nu'}(\omega_n) + \nonumber \\
&+& \frac{1}{4}\sum\limits_{\gamma\gamma'\delta\delta'}{\hat{\cal K}}_{\mu\mu'\gamma\gamma'}(\omega_n){\hat{\cal R}}^{0}_{\gamma\gamma'\delta\delta'}(\omega_n){\hat{\cal T}}_{\delta\delta'\nu\nu'}(\omega_n),\nonumber\\  
\label{Tmatrix}
\eea
i.e., ${\hat{\cal K}}(\omega_n) = {\hat{\cal T}}^{irr}(\omega_n)$. 

Eqs. (\ref{RTmatrixT} -- \ref{Tmatrix}) set the complete and most general framework for the finite-temperature superfluid response in the 4$\times$4 block matrix form and in the domain of Matsubara frequencies, which is remarkably analogous to the non-superfluid and zero-temperature Dyson-BSEs.  Further details depend essentially on the assumptions or direct knowledge about the ground and excited state structure, which can be expressed via expansions of quasiparticle operator strings. Some approximate solutions are discussed in the next subsections.
%\cite{Schuck2019}.

\subsection{Norm matrix, static kernel, and  finite-temperature QRPA (FT-QRPA)}

The norm matrix defined by Eq. (\ref{normT}) reads, explicitly,
\bea
{\hat{\cal N}}_{\mu\mu'\nu\nu'} = \ \ \ \ \ \ \ \ \ \ \ \ \ \ \ \ \ \ \ \ \ \ \ \ \ \ \ \ \ \ \ \nonumber\\
\langle\left(\begin{array}{cccc} [A_{\mu\mu'},A^{\dagger}_{\nu\nu'}] & [A_{\mu\mu'},A_{\nu\nu'}] &  [A_{\mu\mu'},C^{\dagger}_{\nu\nu'}]
&  [A_{\mu\mu'},C_{\nu\nu'}]\\
[3pt]
\left[A^{\dagger}_{\mu\mu'},A^{\dagger}_{\nu\nu'}\right] & [A^{\dagger}_{\mu\mu'},A_{\nu\nu'}] &  [A^{\dagger}_{\mu\mu'},C^{\dagger}_{\nu\nu'}]
&  [A^{\dagger}_{\mu\mu'},C_{\nu\nu'}]\\
[3pt]
\left[C_{\mu\mu'},A^{\dagger}_{\nu\nu'}\right] & [C_{\mu\mu'},A_{\nu\nu'}] &  [C_{\mu\mu'},C^{\dagger}_{\nu\nu'}]
&  [C_{\mu\mu'},C_{\nu\nu'}]\\
[3pt]
\left[C^{\dagger}_{\mu\mu'},A^{\dagger}_{\nu\nu'}\right] & [C^{\dagger}_{\mu\mu'},A_{\nu\nu'}] &  [C^{\dagger}_{\mu\mu'},C^{\dagger}_{\nu\nu'}]
&  [C^{\dagger}_{\mu\mu'},C_{\nu\nu'}]
\end{array}\right)\rangle \nonumber\\
= \langle\left(\begin{array}{cccc}
B_{\mu\mu'\nu\nu'} & 0 & -D_{\nu'\nu\mu\mu'} & -D_{\nu\nu'\mu\mu'} \\
0 & -B^{\dagger}_{\mu\mu'\nu\nu'} & D^{\dagger}_{\nu\nu'\mu\mu'} & D^{\dagger}_{\nu'\nu\mu\mu'} \\
-D^{\dagger}_{\mu'\mu\nu\nu'} & D_{\mu\mu'\nu\nu'} & G_{\mu\mu'\nu\nu'} & G_{\mu\mu'\nu'\nu} \\
-D^{\dagger}_{\mu\mu'\nu\nu'} & D_{\mu'\mu\nu\nu'} & -G^{\dagger}_{\mu\mu'\nu'\nu}  & G^{\dagger}_{\mu\mu'\nu\nu'} 
\end{array}\right)\rangle
\nonumber\\
\eea
and operators $B, D$ and $G$ are listed in Appendix \ref{AppB}.  One can see, in particular, that the main diagonal of the norm matrix is formed by combinations of the $C$ and $C^{\dagger}$ operators, while the non-diagonal matrix elements are either zero (upper left block) or $A$-type operators (upper right and lower left blocks), or $C$ type ones but with one pair of permuted indices (lower right block). In the uncorrelated superfluid quasiparticle states, i.e., the FT-HFB states, all the non-diagonal contributions vanish, and the diagonal matrix elements are ${\cal N}^{[ii]\text {HFB}}_{\mu\mu'\nu\nu'} = \delta_{\mu\mu'\nu\nu'}{\cal N}^{[ii]}_{\mu\mu'}$,
\bea
{\cal N}^{[11]}_{\mu\mu'} &=& 1-n_{\mu}-n_{\mu'}, \ \ \ \ \ \ \ \ \ \ {\cal N}^{[22]}_{\mu\mu'} = - {\cal N}^{[11]}_{\mu\mu'}\nonumber \\
{\cal N}^{[33]}_{\mu\mu'} &=& n_{\mu}-n_{\mu'}, \ \ \ \ \ \ \ \ \ \ \ \ \ \ \  {\cal N}^{[44]}_{\mu\mu'} = - {\cal N}^{[33]}_{\mu\mu'}, \nonumber\\
n_{\mu} &=& \frac{1}{1+\text{exp}(E_{\mu}/T)}.
\label{HFB_norm}
\eea
%The minimally correlated superfluid interaction kernel is confined by the static contribution  
The static contribution to the interaction kernel reads:
\be
{\hat {\cal K}}^{0}_{\mu\mu'\nu\nu'} =  \frac{1}{4}\sum\limits_{\gamma\gamma'\delta\delta'}{\hat {\cal N}}^{-1}_{\mu\mu'\gamma\gamma'}{\hat{\cal T}}^{0}_{\gamma\gamma'\delta\delta'} {\hat{\cal N}}^{-1}_{\delta\delta'\nu\nu'},
\label{K0T}
\ee
where ${\hat{\cal T}}^{0}$ without arguments stands for the matrix of the averaged double commutators
\be
{\hat{\cal T}}^{0}_{\mu\mu'\nu\nu'} =  - \langle  [[V,{\hat\Psi}_{\mu\mu'}],{\hat{\bar\Psi}}_{\nu\nu'}]\rangle .
\label{T0T}
\ee
The minimally correlated superfluid interaction kernel is confined by the static contribution 
and, in combination with the HFB norm matrix, constitutes the FT-QRPA in the Dyson-BSE form:
\bea
{\hat{\cal R}}_{\mu\mu'\nu\nu'}(\omega_n) &=& {\hat{\cal R}}^{0;\text {HFB}}_{\mu\mu'\nu\nu'}(\omega_n) + \nonumber \\
&+& \frac{1}{4}\sum\limits_{\gamma\gamma'\delta\delta'}{\hat{\cal R}}^{0;\text {HFB}}_{\mu\mu'\gamma\gamma'}(\omega_n)\ {\hat{\cal K}}^{0;\text {QRPA}}_{\gamma\gamma'\delta\delta'}\ {\hat{\cal R}}_{\delta\delta'\nu\nu'}(\omega_n),\nonumber\\  
\label{FTQRPA}
\eea
where the index "HFB" indicates the  HFB approximation to the norm (\ref{HFB_norm}) in the uncorrelated propagators, and "QRPA" marks the kernel which, in addition to the norm, uses the uncorrelated HFB variant of the generalized two-body density matrix defined in Appendix \ref{AppE} and the well-known quasi-boson approximation \cite{RingSchuck1980}.
%Setting the ansatz $|i\rangle = Q^{i\dagger}|0\rangle$ with the excitation operator
The finite-temperature grand canonical ensemble averages for the FT-HFB matrix elements of Eq. (\ref{T0T}) can be found using the commutators of Appendix \ref{AppB}, and the results are given in Appendix B of Ref. \cite{Sommermann1983}. The correlated static kernel is presented in Appendix \ref{AppE} in terms of its diagonal components, while the off-diagonal components can be found in a similar fashion.

\subsection{The dynamical kernel}

The most general 4$\times$4 dynamical kernel is given by
\be
{\tilde{\cal K}}^{r[ij]}_{\mu\mu'\nu\nu'}(\tau-\tau') = \langle {\cal T}[V,\Psi^{[i]}_{\mu\mu'}](\tau)[V,{\bar\Psi}^{[j]}_{\nu\nu'}](\tau') 
\rangle^{irr} ,
\label{Krij}
\ee
where the symbol "$\tilde{\ }$" indicates that this is the kernel before including the norm factors, i.e., 
\bea
{\tilde{\cal K}}^{r[ij]}_{\mu\mu'\nu\nu'}(\tau-\tau') &=& {\cal T}^{r[ij]irr}_{\mu\mu'\nu\nu'}(\tau-\tau'), \\
{{\cal K}}^{r[ij]}_{\mu\mu'\nu\nu'}(\tau-\tau') &=&  \frac{1}{4}\sum\limits_{\gamma\gamma'\delta\delta'}{\hat {\cal N}}^{-1}_{\mu\mu'\gamma\gamma'}
{\tilde{\cal K}}^{r[ij]}_{\gamma\gamma'\delta\delta'}(\tau-\tau'){\hat{\cal N}}^{-1}_{\delta\delta'\nu\nu'}. \nonumber \\
\eea
It consists of products of the same set of commutators $[V,A]$, $[V,C]$ given in Appendix \ref{AppB} and their Hermitian conjugates, and has in addition a $\tau$-dependence.
We note that each term in Eq. (\ref{Krij}) contains a product of eight quasiparticle operators, four at time $\tau$ and four at time $\tau'$, i.e., correlated two-times four-quasiparticle propagator, contracted with two matrix elements of the residual interaction $H^{kl}$. As in the non-superfluid and zero-temperature cases, the appearance of higher-rank propagators generates a hierarchy of coupled EOMs for propagators of increasing rank. This means that to find viable solutions for realistic many-body systems, some reasonable approximations have to be applied to the dynamical kernel. Since the generic features of the EOM for the superfluid response, for instance, the interaction kernel decomposition and its operator ranks, are inherited by the finite-temperature theory from the zero-temperature one, we can apply a similar logic as was done, for instance, in Ref. \cite{Litvinova2022}. 

The approximation confined by only one-fermion correlation functions is given by the complete factorization of the ${\cal K}^{r[ij]}_{\mu\mu'\nu\nu'}(\tau-\tau')$ into time-ordered quasiparticle operator pairs. This constitutes the finite-temperature superfluid analog of the second random phase approximation \cite{YannouleasJangChomaz1985}. Correlations that are more relevant to large fermionic systems with intermediate coupling should retain two-fermion correlation functions, which are known to form collective modes. Therefore, in the following, we will focus on such factorization of ${\cal K}^{r[ij]}_{\mu\mu'\nu\nu'}(\tau-\tau')$ into pairwise products of two-quasiparticle propagators. As in Ref. \cite{Litvinova2022} with zero-temperature formalism, mapping the two-quasiparticle correlated pairs to the superfluid phonons, introducing quasiparticle-vibration coupling and
subsequently relaxing correlations in one propagator of each pair leads to the finite-temperature version of the leading-order superfluid NFT. Furthermore, the pairwise factorization enables a closed form of the Dyson-BSE with respect to the response function, with the interesting possibility to iterate the dynamical kernel and, thus, include, in principle, arbitrarily complex 2n-quasiparticle configurations in the factorized form. 

The major difficulty at this step is the complexity of the residual interaction (\ref{Hqua2}) and an even larger number of components than in the case of the zero-temperature superfluid response. However, they have an analogous operator structure. The leading contributions to the dynamical kernel reside on the main diagonal of the $\hat{\cal K}^r$ matrix. The off-diagonal matrix elements represent various 
qPVC-induced ground state correlations of the resonant character. We will discuss the diagonal components in detail, and the off-diagonal ones can be found analogously.
Let us consider the component ${\cal T}^{r[11]}(\tau-\tau')$
and first concentrate on the terms associated with the $H^{31}$ matrix elements, which played the leading role at zero temperature. 
%
% It has been established, in particular, in Ref. \cite{RingSchuck1980} that the terms associated with $H^{40}$ are responsible for the complex ground state %correlations, while the $H^{22}$ contributions have the same operator structure. The leading contributions, thus, come from the $H^{31}$ terms. 
%
As in the latter case, there are three types of operator products at $H^{31}$: (i) '$AA$'-terms, containing operators $A$ and $A^{\dagger}$, (ii)  '$CC$'-terms with only $C$ and  $C^{\dagger}$ operators and (iii) '$AC$'-terms containing both operator types.
The contribution of the first group of terms to ${\cal T}^{r[11]}_{\mu\mu'\nu\nu'}(\tau-\tau')$ reads:
\bea
{\cal T}^{r[11]AA}_{\mu\mu'\nu\nu'}(\tau-\tau') = -\langle {\cal T}\sum\limits_{\rho\rho'\gamma} \Bigl[\bigl(H^{31}_{\rho\rho'\mu\gamma}(A^{\dagger}_{\rho\rho'}A_{\mu'\gamma})(\tau) \nonumber \\ + 
H^{31\ast}_{\rho\rho'\gamma\mu}(A_{\mu'\gamma}A_{\rho\rho'})(\tau)\bigr) - (\mu\leftrightarrow\mu')\Bigr]\nonumber \\ 
\times \sum\limits_{\eta\eta'\delta} \Bigl[\bigl(H^{31\ast}_{\eta\eta'\nu\delta}(A^{\dagger}_{\nu'\delta}A_{\eta\eta'})(\tau') \nonumber \\ + 
H^{31}_{\eta\eta'\delta\nu}(A^{\dagger}_{\eta\eta'}A^{\dagger}_{\nu'\delta})(\tau')\bigr) - (\nu\leftrightarrow\nu')\Bigr] \rangle.
\nonumber \\ 
\eea
The irreducible factorization into $2q$ CFs of the, e.g., the first operator product can be performed as follows:
\bea
\langle &{\cal T}& (A^{\dagger}_{\rho\rho'}A_{\mu'\gamma})(\tau)(A^{\dagger}_{\nu'\delta}A_{\eta\eta'})(\tau')\rangle \nonumber\\
&\approx& 
\langle {\cal T} A^{\dagger}_{\rho\rho'}(\tau)A^{\dagger}_{\nu'\delta}(\tau')\rangle\langle {\cal T}A_{\mu'\gamma}(\tau)A_{\eta\eta'}(\tau')\rangle \nonumber \\ 
&+&
\langle {\cal T} A^{\dagger}_{\rho\rho'}(\tau)A_{\eta\eta'}(\tau')\rangle\langle {\cal T} A_{\mu'\gamma}(\tau)A^{\dagger}_{\nu'\delta}(\tau')\rangle \nonumber \\ 
&+&
\langle {\cal T} C_{\rho\gamma}(\tau)C_{\nu'\eta'}(\tau')\rangle\langle {\cal T} C_{\rho'\mu'}(\tau)C_{\delta\eta}(\tau')\rangle \nonumber\\
&+&
\langle {\cal T} C_{\rho\gamma}(\tau)C_{\delta\eta}(\tau')\rangle\langle {\cal T} C_{\rho'\mu'}(\tau)C_{\nu'\eta'}(\tau')\rangle \nonumber\\
&-&
\langle {\cal T} C_{\rho\gamma}(\tau)A^{\dagger}_{\nu'\delta}(\tau')\rangle\langle {\cal T} C_{\rho'\mu'}(\tau)A_{\eta\eta'}(\tau')\rangle \nonumber\\
&-&
\langle {\cal T} C_{\rho\gamma}(\tau)A_{\eta\eta'}(\tau')\rangle\langle {\cal T} C_{\rho'\mu'}(\tau)A^{\dagger}_{\nu'\delta}(\tau')\rangle \nonumber\\
&-&
\langle {\cal T} A^{\dagger}_{\rho\rho'}(\tau)C_{\nu'\eta'}(\tau')\rangle\langle {\cal T} A_{\mu'\gamma}(\tau)C_{\delta\eta}(\tau')\rangle \nonumber\\
&-&
\langle {\cal T} A^{\dagger}_{\rho\rho'}(\tau)C_{\delta\eta}(\tau')\rangle\langle {\cal T} A_{\mu'\gamma}(\tau)C_{\nu'\eta'}(\tau')\rangle 
- {\cal AS} \nonumber\\
&=& {\cal R}^{[21]}_{\rho\rho'\nu'\delta}(\Delta\tau){\cal R}^{[12]}_{\mu'\gamma\eta\eta'}(\Delta\tau) + {\cal R}^{[22]}_{\rho\rho'\eta\eta'}(\Delta\tau){\cal R}^{[11]}_{\mu'\gamma\nu'\delta}(\Delta\tau) \nonumber\\
&+& {\cal R}^{[34]}_{\rho\gamma\nu'\eta'}(\Delta\tau){\cal R}^{[34]}_{\rho'\mu'\delta\eta}(\Delta\tau) + {\cal R}^{[34]}_{\rho\gamma\delta\eta}(\Delta\tau){\cal R}^{[34]}_{\rho'\mu'\nu'\eta'}(\Delta\tau) \nonumber\\
&-& {\cal R}^{[31]}_{\rho\gamma\nu'\delta}(\Delta\tau){\cal R}^{[32]}_{\rho'\mu'\eta\eta'}(\Delta\tau) - {\cal R}^{[32]}_{\rho\gamma\eta\eta'}(\Delta\tau){\cal R}^{[31]}_{\rho'\mu'\nu'\delta}(\Delta\tau) \nonumber\\
&-& {\cal R}^{[24]}_{\rho\rho'\nu'\eta'}(\Delta\tau){\cal R}^{[14]}_{\mu'\gamma\delta\eta}(\Delta\tau) - {\cal R}^{[23]}_{\rho\rho'\delta\eta}(\Delta\tau){\cal R}^{[13]}_{\mu'\gamma\nu'\eta'}(\Delta\tau) \nonumber\\
&-& {\cal AS},
\label{Tr11AA1}
\eea
where $\Delta\tau = \tau - \tau'$.
The remaining 
%three '$AA$'-terms 
terms can be processed analogously, 
and we retain the ${\cal R}^{[ij]}$ CFs with $[i,j] = [3,4]$ which should also appear in the complete zero-temperature formalism, where they thereby contribute only via the dynamical kernel. These are the new contributions as compared to Ref. \cite{Litvinova2022}. Here we note that ${\cal R}^{[34]}$ components can be expressed via ${\cal R}^{[33]}, {\cal R}^{[43]}$ and ${\cal R}^{[44]}$ as ${\cal U}^{if} = {\cal V}^{ifT}$ by definition (\ref{XYUV}), so that such replacements only lead to index permutations but do not change the CF type.

All the other factorizations are processed in a similar manner. Here we note that the resulting dynamical kernel is composed of terms with the same imaginary time dependence. Thus, the final frequency dependence is also uniform and can be evaluated by the following transformation: 
\bea
[{\cal R}^{[ij]}_{\mu\mu'\nu\nu'}{\cal R}^{[kl]}_{\eta\eta'\rho\rho'}](\omega_k) = \int_{-1/T}^{1/T}d\tau e^{\omega_k\tau}{\cal R}^{[ij]}_{\mu\mu'\nu\nu'}(\tau){\cal R}^{[kl]}_{\eta\eta'\rho\rho'}(\tau) \nonumber\\
= \sum\limits_{ifi'f'}p_ip_{i'}\sum\limits_{\sigma = \pm}\sigma\frac{{\cal Z}^{if(i\sigma)}_{\mu\mu'}{\cal Z}^{if(j\sigma)\ast}_{\nu\nu'}
{\cal Z}^{i'f'(k\sigma)}_{\eta\eta'}{\cal Z}^{i'f'(l\sigma)\ast}_{\rho\rho'}}
{\omega_k - \sigma(\omega_{fi} + \omega_{f'i'})}\nonumber\\
\times\bigl(e^{-(\omega_{fi} + \omega_{f'i'})/T} - 1\bigr), \ \ \ \ \ \ \ \ \ \ \ \ \ \ 
\label{FTRR}
\eea
where
$\omega_{fi} = E_{f} - E_{i}$.
Contracting the factorized CFs with the interaction matrix elements $H^{31}$, $H^{40}$, and $H^{22}$, one finds that, as in the zero-temperature case, these contractions form two types which were classified as types (a) and  (b) in Ref. \cite{Litvinova2022}. A similar classification can be made at finite temperature, where only the number of terms grows because of the additional contributions from the $\{{\cal U}, {\cal V}\}$ amplitudes. In particular, in Eq. (\ref{Tr11AA1}), the first terms in the last four lines are of type (b) and the second terms in those lines are of type (a). 
Collecting all the terms of type (a) in the dynamical kernel and dropping the reducible terms belonging to the constant and one-body sectors of the Hamiltonian, the compact form of $\tilde{\cal K}^{r[11]}$ and  $\tilde{\cal K}^{r[33]}$ (again, before the inclusion of the norm factors) in the energy domain can be recast as follows:
%
%%%%%%%%%%%%%% Before adding cross terms
%\bea
%{\cal K}^{r[11]cc(a)}_{\mu\mu'\nu\nu'}(\omega_k) = 
%\sum\limits_{\gamma\delta nm}p_{i_n}p_{i_m}
%\Bigl[\frac{\Gamma^{(11)n}_{\mu\gamma}{\cal X}^{m}_{\mu'\gamma}{\cal X}^{m\ast}_{\nu'\delta}\Gamma^{(11)n\ast}_{\nu\delta}}{\omega_k - \omega_n - \omega_m}  \nonumber\\ - 
%\frac{\Gamma^{(11)n\ast}_{\gamma\mu}{\cal Y}^{m\ast}_{\mu'\gamma}{\cal Y}^{m}_{\nu'\delta}\Gamma^{(11)n}_{\delta\nu}}{\omega_k + \omega_n + \omega_m}
%+ \frac{\Gamma^{(02)n}_{\mu\gamma}{\cal U}^{m}_{\mu'\gamma}{\cal U}^{m\ast}_{\nu'\delta}\Gamma^{(02)n\ast}_{\nu\delta}}{\omega_k - \omega_n - \omega_m} \nonumber\\ - 
%\frac{\Gamma^{(02)n\ast}_{\gamma\mu}{\cal V}^{m\ast}_{\mu'\gamma}{\cal V}^{m}_{\nu'\delta}\Gamma^{(02)n}_{\delta\nu}}{\omega_k + \omega_n + \omega_m}
%\Bigr]\bigl(e^{-(\omega_{n} + \omega_{m})/T} - 1\bigr)
%- \cal{AS},\nonumber\\
%\label{Kr11cc}
%\eea
%%%%%%%%%%%%%% With cross terms
\bea
%&\ &
{\tilde{\cal K}}^{r[11]cc(a)}_{\mu\mu'\nu\nu'}(\omega_k) = \sum_{\gamma \delta n m}p_{i_{n}}p_{i_{m}} (1 - e^{-(\omega_n + \omega_m)/T})\nonumber\\ 
%\bigl[ %\frac{\langle [V, A_{\mu\mu'}] \rangle^{nm} \langle [V, A^{\dagger}_{\nu\nu'}] %\rangle^{nm \ast}}{\omega_k - \omega_n -\omega_m} - \frac{\langle [V, A_{\mu\mu'}] %\rangle^{\bar{n}\bar{m}} \langle [V, %A^{\dagger}_{\nu\nu'}] \rangle^{\bar{n}\bar{m} %\ast}}{\omega_k + \omega_n %+\omega_m} \bigr] = 
%\sum\limits_{\gamma\delta nm}p_{i_n}p_{i_m}\bigl(1 - e^{-(\omega_n + %\omega_m)/T}\bigr)\nonumber \\
\times\Bigl[\frac{(\Gamma^{(11)n}_{\mu\gamma}{\cal X}^{m}_{\mu'\gamma} + \Gamma^{(20)n}_{\mu\gamma}{\cal V}^{m}_{\mu'\gamma})
({\cal X}^{m\ast}_{\nu'\delta}\Gamma^{(11)n\ast}_{\nu\delta} + {\cal V}^{m\ast}_{\nu'\delta}\Gamma^{(20)n\ast}_{\nu\delta})}{\omega_k - \omega_n - \omega_m}  
\nonumber\\ - 
\frac{(\Gamma^{(11){\bar n}}_{\mu\gamma}{\cal X}^{\bar m}_{\mu'\gamma} + \Gamma^{(20){\bar n}}_{\mu\gamma}{\cal V}^{\bar m}_{\mu'\gamma})
({\cal X}^{{\bar m}\ast}_{\nu'\delta}\Gamma^{(11){\bar n}\ast}_{\nu\delta} + {\cal V}^{{\bar m}\ast}_{\nu'\delta}\Gamma^{(20){\bar n}\ast}_{\nu\delta})}{\omega_k + \omega_n + \omega_m}\Bigr]\nonumber\\
%
%+ \frac{\Gamma^{(02)n}_{\mu\gamma}{\cal U}^{m}_{\mu'\gamma}{\cal U}^{m\ast}_{\nu'\delta}\Gamma^{(02)n\ast}_{\nu\delta}}{\omega_k - \omega_n - \omega_m} \nonumber\\ - 
%\frac{\Gamma^{(02)n\ast}_{\gamma\mu}{\cal V}^{m\ast}_{\mu'\gamma}{\cal V}^{m}_{\nu'\delta}\Gamma^{(02)n}_{\delta\nu}}{\omega_k + \omega_n + \omega_m}
%\Bigr]\bigl(e^{-(\omega_{n} + \omega_{m})/T} - 1\bigr)
- \cal{AS}, \ \ \ \ \ \ \  
\label{Kr11cc}
\eea

\bea
%&\ &
{\tilde{\cal K}}^{r[33]cc(a)}_{\mu\mu'\nu\nu'}(\omega_k) = \sum_{\gamma \delta n m}p_{i_{n}}p_{i_{m}} (1 - e^{-(\omega_n + \omega_m)/T})\nonumber\\ 
\times\Bigl[\frac{(\Gamma^{(11)n}_{\gamma \mu}{\cal U}^{m}_{\gamma \mu'} + \Gamma^{(20)n}_{\mu'\gamma}{\cal Y}^{m}_{\mu\gamma})
({\cal U}^{m\ast}_{\delta\nu'}\Gamma^{(11)n\ast}_{\delta\nu} + {\cal Y}^{m\ast}_{\nu\delta}\Gamma^{(20)n\ast}_{\nu'\delta})}{\omega_k - \omega_n - \omega_m}  
\nonumber\\ - 
\frac{(\Gamma^{(11){\bar n}}_{\gamma\mu}{\cal U}^{\bar m}_{\gamma\mu'} + \Gamma^{(20){\bar n}}_{\mu'\gamma}{\cal Y}^{\bar m}_{\mu\gamma})
({\cal U}^{{\bar m}\ast}_{\delta\nu'}\Gamma^{(11){\bar n}\ast}_{\delta\nu} + {\cal Y}^{{\bar m}\ast}_{\nu\delta}\Gamma^{(20){\bar n}\ast}_{\nu'\delta})}{\omega_k + \omega_n + \omega_m}\Bigr]\nonumber\\
- \cal{AS}^{\dagger}, \ \ \ \ \ \  \\
% alternative
= \sum_{\gamma \delta n m}p_{i_{n}}p_{i_{m}} (1 - e^{-(\omega_n + \omega_m)/T})\nonumber\\ 
\times\Bigl[\frac{(\Gamma^{(11){\bar n}\ast}_{\mu\gamma}{\cal U}^{{\bar m}\ast}_{\mu'\gamma } - \Gamma^{(20){\bar n}\ast}_{\mu\gamma}{\cal Y}^{{\bar m}\ast}_{\mu'\gamma})
({\cal U}^{{\bar m}}_{\nu'\delta}\Gamma^{(11){\bar n}}_{\nu\delta} - {\cal Y}^{{\bar m}}_{\nu'\delta}\Gamma^{(20){\bar n}}_{\nu\delta})}{\omega_k - \omega_n - \omega_m}  
\nonumber\\ - 
\frac{(\Gamma^{(11){n}\ast}_{\mu\gamma}{\cal U}^{{m}\ast}_{\mu'\gamma } - \Gamma^{(20){n}\ast}_{\mu\gamma}{\cal Y}^{{m}\ast}_{\mu'\gamma})
({\cal U}^{{m}}_{\nu'\delta}\Gamma^{(11){n}}_{\nu\delta} - {\cal Y}^{{m}}_{\nu'\delta}\Gamma^{(20){n}}_{\nu\delta})}{\omega_k + \omega_n + \omega_m}\Bigr]\nonumber\\
- \cal{AS}^{\dagger}, \ \ \ \ \ \ 
\label{Kr33cc}
\eea
where $\cal{AS}$ stands for the usual antisymmetrization $(\mu\leftrightarrow \mu')$ and $(\nu\leftrightarrow \nu')$, while $\cal{AS}^{\dagger}$ is defined in Appendix \ref{AppD}. 
The multi-indices $n = \{i_nf_n\}$ and $m = \{i_mf_m\}$ are introduced to simplify the notations, ${\bar n} = \{f_ni_n\}$, and the superfluid phonon vertices completed by the $\{{\cal U}, {\cal V}\}$-amplitudes are specified in Appendix \ref{AppC}. As in Ref. \cite{Litvinova2022}, the additional index "cc" marks the factorization with two correlation functions. The new features, as compared to the zero-temperature case, are the distributions $p_{i_n}$ and the exponential factors, in addition to the double summations over the initial states. Besides that, one finds the new terms with the $\{{\cal U}, {\cal V}\}$ amplitudes, which appear with the $\Gamma^{(02)n}$ vertices, representing the diagonal part of the qPVC-associated ground state correlations. The ${\tilde{\cal K}}^{r[11]cc}_{\mu\mu'\nu\nu'}(\omega_k)$ of Eq. (\ref{Kr11cc}) can be used as the reference component for the entire dynamical kernel, so that the other components can be found analogously. For instance, in the ${\tilde{\cal K}}^{r[33]cc}_{\mu\mu'\nu\nu'}(\omega_k)$ matrix elements the amplitudes $\{{\cal X},{\cal Y}\}$ and $\{{\cal U},{\cal V}\}$ switch their roles as compared to the ${\tilde{\cal K}}^{r[11]cc}_{\mu\mu'\nu\nu'}(\omega_k)$ one, while ${\tilde{\cal K}}^{r[22]}$ and ${\tilde{\cal K}}^{r[44]}$ are related, respectively, to  ${\tilde{\cal K}}^{r[11]}$ and ${\tilde{\cal K}}^{r[33]}$ by Hermitian conjugation. We note that the residues in Eq. (\ref{Kr11cc}) are nicely factorized so that block-non-diagonal matrix elements of the dynamical kernel can be found by recombining the amplitudes $\{{\cal X},{\cal Y},{\cal U},{\cal V}\}$ and vertices $\{\Gamma^{(11)},\Gamma^{(02)}\}$ according to their operator structure determined by Eq. (\ref{Krij}). Thus, generalizing Eqs. (\ref{Kr11cc},\ref{Kr33cc}), one finds
\bea
{\tilde{\cal K}}^{r[ij]cc(a)}_{\mu\mu'\nu\nu'}(\omega_k) &=& \sum_{\gamma \delta n m}p_{i_{n}}p_{i_{m}} (1 - e^{-(\omega_n + \omega_m)/T})\nonumber\\
&\times&\Bigl[ \frac{\langle [V, \hat{\Psi}^{[i]}_{\mu\mu'}] \rangle^{nm} \langle [V, \hat{\Psi}^{[j]}_{\nu\nu'}] \rangle^{nm \ast}}{\omega_k - \omega_n -\omega_m} 
\nonumber\\
&-& \frac{\langle [V, \hat{\Psi}^{[i]}_{\mu\mu'}] \rangle^{\bar{n}\bar{m}} \langle [V, \hat{\Psi}^{[j]}_{\nu\nu'}] \rangle^{\bar{n}\bar{m}\ast}}{\omega_k + \omega_n +\omega_m} \Bigr],
%- \cal{AS}, \ \ \ \ \ \ 
\nonumber\\
\label{Krijcca}
\eea
where the expectation values $\langle [V, \hat{\Psi}^{[i]}_{\mu\mu'}] \rangle^{nm}$ are specified in Appendix \ref{AppD}.
Furthermore, each type (a) component has a type (b) counterpart, where, in each residue, the indices $m$ and $n$ in the second pair of the transition amplitude and qPVC vertex interchange \cite{Litvinova2022}:
\bea
{\tilde{\cal K}}^{r[ij]cc(b)}_{\mu\mu'\nu\nu'}(\omega_k) &=& \sum_{\gamma \delta n m}p_{i_{n}}p_{i_{m}} (1 - e^{-(\omega_n + \omega_m)/T})\nonumber\\
&\times&\Bigl[ \frac{\langle [V, \hat{\Psi}^{[i]}_{\mu\mu'}] \rangle^{nm} \langle [V, \hat{\Psi}^{[j]}_{\nu\nu'}] \rangle^{mn \ast}}{\omega_k - \omega_n -\omega_m} 
\nonumber\\
&-& \frac{\langle [V, \hat{\Psi}^{[i]}_{\mu\mu'}] \rangle^{\bar{n}\bar{m}} \langle [V, \hat{\Psi}^{[j]}_{\nu\nu'}] \rangle^{\bar{m}\bar{n}\ast}}{\omega_k + \omega_n +\omega_m} \Bigr].
%- \cal{AS}, \ \ \ \ \ \ 
\label{Krijccb}
\eea
Thereby, Eqs. (\ref{Kr11cc} - \ref{Krijccb}) represent the complete dynamical kernel in the "cc"-factorization scheme. 
This factorization enables a closed EOM for the response function in terms of the two-quasiparticle amplitudes $\{{\cal X},{\cal Y}\}$ and $\{{\cal U},{\cal V}\}$, which can be solved by an iterative procedure, similarly to the non-superfluid case of Ref. \cite{LitvinovaSchuck2019}, confined by  $\{{\cal X},{\cal Y}\}$. 
Strictly speaking, to have a fully closed matrix EOM with the bare interaction as the only input, the static kernel (\ref{T0sym} - \ref{T033}) should also be factorized. 
%%%%%%

Remarkably, here all the dynamical ground state correlations are included via the amplitudes $\{{\cal U},{\cal V}\}$ and off-diagonal components. The zero-temperature limit of Eqs. (\ref{Kr11cc} - \ref{Krijccb}) can be easily taken, which means that the $T=0$ superfluid theory is also completed. Numerical implementations of the response theory with the dynamical kernels derived in this section will be explored for both $T=0$ and $T>0$ in future work. 

\section{Summary and outlook}
\label{summary}
We presented a response theory for superfluid fermionic systems at finite temperatures. The theory is formulated in the QFT language for fermions interacting via unspecified bare two-body forces. It starts with a definition of the two-quasiparticle superfluid thermal response function in the frequency domain, compatible with the structural part of the spectral strength distribution and with the two-time two-fermion Matsubara propagator in a correlated medium, averaged over the grand canonical ensemble in the imaginary time domain. The latter form enables a derivation of the most general EOM for this $2q$ propagator. 
As feedback, accounting for both forward and backward components of the response requires an additional thermal prefactor in the expression for the spectral strength distribution as compared to the zero-temperature case, which changes the low-energy limit of the strength distribution because of a zero-energy singularity. 

The complete thermal superfluid formalism requires an extension of the regular zero-temperature 2$\times$2 component structure characterizing the fermionic response to the 4$\times$4 arrays. This is dictated by the presence of transition amplitudes of the ${\cal U}_{\mu\mu'} = \alpha^{\dagger}_{\mu}\alpha_{\mu'}$ type, in addition to the commonly taken into account $\{{\cal X},{\cal Y}\}$ ones in QRPA and its extensions. Therefore, we included contributions from the ${\cal U}$-type amplitudes, which vanish in QRPA at zero temperature, but acquire non-zero components in the norm matrix at finite temperature and thus become a non-negligible part of the EOM. This is a well-known feature of the FT-QRPA \cite{Sommermann1983}, which in this work is obtained in the FT-HFB limit of the static kernel and with the completely neglected dynamical kernel of the $2q$ EOM.

The ${\cal U}$-type amplitudes contribute non-trivially to the dynamical kernel of the superfluid finite-temperature $2q$ EOM, i.e., in the FT-QRPA extension. Besides the formal extension of this kernel to the 4$\times$4 block form, these amplitudes further complicate the phonon vertex composition, appearing in the factorization approximations to the dynamical kernel which can be mapped to qPVC. Another new feature of the dynamical kernel is its explicit temperature dependence, in addition to that in the statistical thermal averages of all the expectation values, also present in the static kernel. 

This work thereby extends the applicability of the finite-temperature non-superfluid response theory formulated in Refs. \cite{LitvinovaWibowo2018,WibowoLitvinova2019} and applied to various nuclear spectra characterizing both neutral and charge-exchange responses \cite{LitvinovaWibowo2018,WibowoLitvinova2019,LitvinovaRobinWibowo2020,Litvinova2021b}. The major new capability of the superfluid finite-temperature theory is bridging the superfluid phase at $T = 0$ and the non-superfluid one at $T\geq T_c$, where $T_c$ is the critical temperature of the superfluid phase transition. The range $0 \leq T \leq T_c$ is the most relevant temperature interval for nuclei undergoing rapid nucleosynthesis in stellar environments and for electron capture in core-collapse supernovae. Extending computation of these phenomena across the $0 \leq T \leq T_c$ temperature regime is thereby one of the most interesting applications of the developed formalism to nuclear structure. Because of its generality, the theory can also be applied across condensed matter phenomenology.

%===============================================================================
\section*{Acknowledgement}
This work was supported by the US-NSF Grants PHY-2209376 and PHY-2515056.
%
%===============================================================================

\appendix
\section{Hamiltonian matrix elements in the quasiparticle space}
\label{AppA}
The matrix elements of the quasiparticle operator part of the fermionic Hamiltonian (\ref{Hqua1}) read \cite{RingSchuck1980}:
\bea
%H^0 &=& \sum\limits_{12}h_{12}\rho_{21} + \frac{1}{2}\sum\limits_{1234}\rho_{31}{\bar v}_{1234}\rho_{42} \nonumber \\
%&+& \frac{1}{4}\sum\limits_{1234}\varkappa^{\ast}_{12}{\bar v}_{1234}\varkappa_{34} 
%\\
H^{11}_{\mu\nu} &=& \sum\limits_{12}\bigl(U^{\dagger}_{\mu 1}h_{12}U_{2\nu} - V^{\dagger}_{\mu 1}h^T_{12}V_{2\nu} 
 \nonumber \\ &+& U^{\dagger}_{\mu 1}\Delta_{12} V_{2\nu} 
- V^{\dagger}_{\mu 1}\Delta^{\ast}_{12}U_{2\nu}\bigr)
\\
H^{20}_{\mu\nu} &=& \sum\limits_{12}\bigl(U^{\dagger}_{\mu 1}h_{12}V^{\ast}_{2\nu} - V^{\dagger}_{\mu 1}h^T_{12}U^{\ast}_{2\nu} 
\nonumber \\ &+& U^{\dagger}_{\mu 1}\Delta_{12} U^{\ast}_{2\nu}  - V^{\dagger}_{\mu 1}\Delta^{\ast}_{12}V^{\ast}_{2\nu}\bigr)
\\
H^{40}_{\mu\mu'\nu\nu'} &=& \frac{1}{4}\sum\limits_{1234}{\bar v}_{1234} U^{\ast}_{1\mu}U^{\ast}_{2\mu'}V^{\ast}_{4\nu}V^{\ast}_{3\nu'}
\\
H^{31}_{\mu\mu'\nu\nu'} &=& \frac{1}{2}\sum\limits_{1234}{\bar v}_{1234}\bigl(U^{\ast}_{1\mu}V^{\ast}_{4\mu'}V^{\ast}_{3\nu}V_{2\nu'} %\nonumber\\
+ V^{\ast}_{3\mu}U^{\ast}_{2\mu'}U^{\ast}_{1\nu}U_{4\nu'}\bigr) \nonumber\\
\label{H31}
\\
H^{22}_{\mu\mu'\nu\nu'} &=& \sum\limits_{1234}{\bar v}_{1234}\Bigl[\bigl(U^{\ast}_{1\mu}V^{\ast}_{4\mu'}V_{2\nu}U_{3\nu'}
- (\mu \to \mu')\bigr)  \nonumber\\  &-& \bigl(\nu \to \nu'\bigr) 
+ U^{\ast}_{1\mu}U^{\ast}_{2\mu'}U_{3\nu}U_{4\nu'} + V^{\ast}_{3\mu}V^{\ast}_{4\mu'}V_{1\nu}V_{2\nu'}
\Bigr].   \nonumber\\
\eea

\section{Generic commutators}
\label{AppB}
%%% Norm
The basic commutators between the different quasiparticle pair operators defining the norm matrix are the following:
\bea
[A_{\mu\mu'},A^{\dagger}_{\nu\nu'}] &=& -\bigl\{[(\delta_{\mu\nu}C_{\nu'\mu'}) - (\mu\leftrightarrow\mu')] - [(\nu\leftrightarrow\nu')]\bigr\} +\nonumber\\
+ \delta_{\mu\mu'\nu\nu'} &=& B_{\mu\mu'\nu\nu'} = B^{\dagger}_{\nu\nu'\mu\mu'} = -B_{\mu'\mu\nu\nu'} = -B_{\mu\mu'\nu'\nu}\nonumber\\
\left[A_{\mu\mu'},C_{\nu\nu'}\right] &=& \delta_{\mu'\nu}A_{\mu\nu'} - \delta_{\mu\nu}A_{\mu'\nu'}  
= D_{\nu\nu'\mu\mu'}\nonumber\\
\left[C_{\mu\mu'},C^{\dagger}_{\nu\nu'}\right] &=& \delta_{\mu'\nu'}C_{\mu\nu} - \delta_{\mu\nu}C^{\dagger}_{\mu'\nu'} = G_{\mu\mu'\nu\nu'}
\eea
%%% Interaction
The basic A-block (upper left part of the kernel matrix) commutators relevant to the interaction kernel read:
\bea
[A^{\dagger}_{\nu\nu'}A^{\dagger}_{\gamma\gamma'},A_{\mu\mu'}] \equiv
[\alpha^{\dagger}_{\nu}\alpha^{\dagger}_{\nu'}\alpha^{\dagger}_{\gamma}\alpha^{\dagger}_{\gamma'},\alpha_{\mu'}\alpha_{\mu}]  
 \nonumber\\
= -B_{\mu\mu'\nu\nu'}A^{\dagger}_{\gamma\gamma'} - A^{\dagger}_{\nu\nu'}B_{\mu\mu'\gamma\gamma'},
\eea
at the non-vanishing contribution of $H^{40}$. The term containing $H^{31}$ is defined by
\bea
[A^{\dagger}_{\nu\nu'}C_{\gamma\gamma'},A_{\mu\mu'}] \equiv [A^{\dagger}_{\nu\nu'}\alpha^{\dagger}_{\gamma}\alpha_{\gamma'},A_{\mu\mu'}] 
\nonumber\\
= A^{\dagger}_{\nu\nu'}D_{\gamma\gamma'\mu\mu'} - B_{\mu\mu'\nu\nu'}C_{\gamma\gamma'},
\eea
\bea
[C_{\gamma'\gamma}A_{\nu\nu'},A_{\mu\mu'}] \equiv [\alpha^{\dagger}_{\gamma'}\alpha_{\gamma}A_{\nu\nu'},A_{\mu\mu'}] \nonumber\\
= D_{\gamma'\gamma\mu\mu'}A_{\nu\nu'},
\eea
\bea
[A^{\dagger}_{\nu\nu'}C_{\gamma\gamma'},A^{\dagger}_{\mu\mu'}] \equiv 
 [A^{\dagger}_{\nu\nu'}\alpha^{\dagger}_{\gamma}\alpha_{\gamma'},A^{\dagger}_{\mu\mu'}] \nonumber\\
=  -A^{\dagger}_{\nu\nu'}D^{\dagger}_{\gamma'\gamma\mu\mu'}.
\eea
\bea
[C_{\gamma'\gamma}A_{\nu\nu'},A^{\dagger}_{\mu\mu'}] \equiv  [\alpha^{\dagger}_{\gamma'}\alpha_{\gamma}A_{\nu\nu'},A^{\dagger}_{\mu\mu'}] 
\nonumber\\
= C_{\gamma'\gamma}B^{\dagger}_{\mu\mu'\nu\nu'} -D^{\dagger}_{\gamma\gamma'\mu\mu'}A_{\nu\nu'},
\eea
and, the commutators associated with $H^{22}$ read:
\bea
[A^{\dagger}_{\gamma\gamma'}A_{\nu\nu'},A_{\mu\mu'}] = -B_{\mu\mu'\gamma\gamma'}A_{\nu\nu'},
\nonumber
\\
\left[A^{\dagger}_{\gamma\gamma'}A_{\nu\nu'},A^{\dagger}_{\mu\mu'}\right] = A^{\dagger}_{\gamma\gamma'}B_{\nu\nu'\mu\mu'}.
\eea
Collecting all the contributions, the reference commutator for the A-block is given by:
\bea
[V,A_{\mu\mu'}] = -\sum\limits_{\nu\nu'\gamma\gamma'} H^{40}_{\nu\nu'\gamma\gamma'} (B_{\mu\mu'\nu\nu'}A^{\dagger}_{\gamma\gamma'} + 
A^{\dagger}_{\nu\nu'}B_{\mu\mu'\gamma\gamma'}) \nonumber\\
+ \sum\limits_{\nu\nu'\gamma\gamma'} \bigl(H^{31}_{\nu\nu'\gamma\gamma'}(A^{\dagger}_{\nu\nu'}D_{\gamma\gamma'\mu\mu'} - B_{\mu\mu'\nu\nu'}C_{\gamma\gamma'}) \nonumber\\
+  H^{31\ast}_{\nu\nu'\gamma\gamma'}D_{\gamma'\gamma\mu\mu'}A_{\nu\nu'}\bigr) \nonumber\\
- \frac{1}{8}\sum\limits_{\nu\nu'\gamma\gamma'} \bigl(H^{22}_{\nu\nu'\gamma\gamma'}B_{\mu\mu'\nu\nu'}A_{\gamma\gamma'} + H^{22\ast}_{\nu\nu'\gamma\gamma'}B_{\mu\mu'\gamma\gamma'}A_{\nu\nu'}\bigr),\nonumber\\
\label{VAcomm}
\eea
while the commutator with $A^{\dagger}$ can be related to the one above from it by Hermitian conjugation:
\be
[V,A^{\dagger}_{\mu\mu'}] = -[V,A_{\mu\mu'}]^{\dagger},
\label{VAdcomm}
\ee
since $V = V^{\dagger}$, i.e., is Hermitian.

The C-block commutators are evaluated analogously:
\bea
[A^{\dagger}_{\nu\nu'}A^{\dagger}_{\gamma\gamma'},C_{\mu\mu'}] \equiv
[\alpha^{\dagger}_{\nu}\alpha^{\dagger}_{\nu'}\alpha^{\dagger}_{\gamma}\alpha^{\dagger}_{\gamma'},\alpha^{\dagger}_{\mu}\alpha_{\mu'}]  
 \nonumber\\
= D^{\dagger}_{\mu'\mu\nu\nu'}A^{\dagger}_{\gamma\gamma'} + A^{\dagger}_{\nu\nu'}D^{\dagger}_{\mu'\mu\gamma\gamma'}
\eea
at $H^{40}$, 
\bea
[A^{\dagger}_{\nu\nu'}C_{\gamma\gamma'},C_{\mu\mu'}] \equiv [A^{\dagger}_{\nu\nu'}\alpha^{\dagger}_{\gamma}\alpha_{\gamma'},C_{\mu\mu'}] 
\nonumber\\
= A^{\dagger}_{\nu\nu'}G_{\gamma\gamma'\mu'\mu} + D^{\dagger}_{\mu'\mu\nu\nu'}C_{\gamma\gamma'},
\eea
\bea
[C_{\gamma'\gamma}A_{\nu\nu'},C_{\mu\mu'}] \equiv [\alpha^{\dagger}_{\gamma'}\alpha_{\gamma}A_{\nu\nu'},C_{\mu\mu'}] \nonumber\\
= -C^{\dagger}_{\gamma\gamma'}D_{\mu\mu'\nu\nu'} - G_{\mu\mu'\gamma\gamma'}A_{\nu\nu'},
\eea
\bea
[A^{\dagger}_{\nu\nu'}C_{\gamma\gamma'},C^{\dagger}_{\mu\mu'}] \equiv [A^{\dagger}_{\nu\nu'}\alpha^{\dagger}_{\gamma}\alpha_{\gamma'},C^{\dagger}_{\mu\mu'}] 
\nonumber\\
= A^{\dagger}_{\nu\nu'}G_{\gamma\gamma'\mu\mu'} + D^{\dagger}_{\mu\mu'\nu\nu'}C_{\gamma\gamma'},
\eea
\bea
[C_{\gamma'\gamma}A_{\nu\nu'},C^{\dagger}_{\mu\mu'}] \equiv  [\alpha^{\dagger}_{\gamma'}\alpha_{\gamma}A_{\nu\nu'},C^{\dagger}_{\mu\mu'}] 
\nonumber\\
= -C^{\dagger}_{\gamma\gamma'}D_{\mu'\mu\nu\nu'} - G^{\dagger}_{\gamma\gamma'\mu'\mu}A_{\nu\nu'}
\eea
at $H^{31}$, and 
\bea
[A^{\dagger}_{\gamma\gamma'}A_{\nu\nu'},C_{\mu\mu'}] = -A^{\dagger}_{\gamma\gamma'}D_{\mu\mu'\nu\nu'} + 
D^{\dagger}_{\mu'\mu\gamma\gamma'}A_{\nu\nu'},
\nonumber
\\
\left[A^{\dagger}_{\gamma\gamma'}A_{\nu\nu'},C^{\dagger}_{\mu\mu'}\right] = -A^{\dagger}_{\gamma\gamma'}D_{\mu'\mu\nu\nu'} + 
D^{\dagger}_{\mu\mu'\gamma\gamma'}A_{\nu\nu'}.\nonumber\\
\eea
at $H^{22}$, so that  
\bea
[V,C_{\mu\mu'}] = \sum\limits_{\nu\nu'\gamma\gamma'} \bigl(H^{40}_{\nu\nu'\gamma\gamma'} (A^{\dagger}_{\nu\nu'}D^{\dagger}_{\mu'\mu\gamma\gamma'} + D^{\dagger}_{\mu'\mu\nu\nu'}A^{\dagger}_{\gamma\gamma'} ) \nonumber \\
- H^{40\ast}_{\nu\nu'\gamma\gamma'} (A_{\gamma\gamma'}D_{\mu\mu'\nu\nu'} + D_{\mu\mu'\gamma\gamma'}A_{\nu\nu'} )\bigr) \nonumber\\
+ \sum\limits_{\nu\nu'\gamma\gamma'} \bigl(H^{31}_{\nu\nu'\gamma\gamma'}(A^{\dagger}_{\nu\nu'}G_{\gamma\gamma'\mu'\mu} + D^{\dagger}_{\mu'\mu\nu\nu'}C_{\gamma\gamma'}) \nonumber\\
-  H^{31\ast}_{\nu\nu'\gamma\gamma'}(G_{\mu\mu'\gamma\gamma'}A_{\nu\nu'} + C^{\dagger}_{\gamma\gamma'}D_{\mu\mu'\nu\nu'})\bigr) \nonumber\\
+ \frac{1}{8}\sum\limits_{\nu\nu'\gamma\gamma'} \bigl(H^{22}_{\nu\nu'\gamma\gamma'}(D^{\dagger}_{\mu'\mu\nu\nu'}A_{\gamma\gamma'} - 
A^{\dagger}_{\nu\nu'}D_{\mu\mu'\gamma\gamma'}) \nonumber\\
+ H^{22\ast}_{\nu\nu'\gamma\gamma'}(D^{\dagger}_{\mu'\mu\gamma\gamma'}A_{\nu\nu'} - 
A^{\dagger}_{\gamma\gamma'}D_{\mu\mu'\nu\nu'})
\bigr).\nonumber\\
\label{VCcomm}
\eea
Similar to the A-block,
\be
[V,C^{\dagger}_{\mu\mu'}] = -[V,C_{\mu\mu'}]^{\dagger}.
\label{VCdcomm}
\ee
\\

\section{Phonon Vertices}
\label{AppC}
The vertices of the superfluid phonons $\Gamma^{(ij)n}$, in the HFB approximation to the intermediate fermionic line, are introduced in analogy to those at zero temperature: 
\bea
\Gamma^{(11)n}_{\mu\nu} 
= 
\Bigl[ 
U^{\dagger}g^{n}U &+& U^{\dagger}\gamma^{n(+)}V 
- V^{\dagger}g^{nT}V 
\nonumber \\
&-& V^{\dagger}\gamma^{n(-)T}U\Bigr]_{\mu\nu}, 
\label{Gamma11_HFB}
\eea
\bea
\Gamma^{(11)n}_{\mu\nu} = \Gamma^{(11)\bar{n}\ast}_{\nu\mu}
\eea
\bea
\Gamma^{(02)n}_{\mu\nu} = 
-\Bigl[ 
V^{T}g^{n}U &+& V^{T}\gamma^{n(+)}V 
- U^{T}g^{nT}V \nonumber \\
&-& U^{T}\gamma^{n(-)T}U\Bigr]_{\mu\nu},
\label{Gamma02_HFB}
\eea
\bea
\Gamma^{(20)n}_{\mu\nu} = 
[ 
U^{\dagger}g^{n}V^{\ast} &+& U^{\dagger}\gamma^{n(+)}U^{\ast} 
- V^{\dagger}g^{nT}U^{\ast} \nonumber \\
&-& V^{\dagger}\gamma^{n(-)T}V^{\ast}\Bigr]_{\mu\nu},
\label{Gamma02_HFB}
\eea
\bea
\Gamma^{(20)n}_{\mu\nu} = -\Gamma^{(20)n}_{\nu\mu}=\Gamma^{(02)\bar{n}\ast}_{\mu\nu}=-\Gamma^{(02)\bar{n}\ast}_{\nu\mu}, 
\eea
where $g^n$ and $\gamma^n$ are the vertices of normal and pairing phonons in the canonical, or single-particle, basis, and $n = \{if\}$ is a multi-index of the initial and final states. 
The vertices $g^n$ and $\gamma^n$ are defined via the normal and anomalous transition densities 
\bea
\rho^{if}_{12} &=& \langle i|\psi^{\dagger}_2\psi_1|f \rangle \nonumber\\
\varkappa^{if(+)}_{12} &=& \langle i|\psi_2\psi_1|f\rangle \nonumber\\
 \varkappa^{if(-)\ast}_{21} &=& \langle i|\psi^{\dagger}_2\psi^{\dagger}_1|f\rangle ,
\label{rho}
\eea
related to the superfluid amplitudes as follows:
\bea
\rho^n_{12} &=&  (U{\cal X}^nV^T + V^{\ast}{\cal Y}^{nT}U^{\dagger}- V^{\ast}{\cal U}^{n}V^{T} + U{\cal V}^{n}U^{\dagger} )_{12}\nonumber\\
\varkappa^{n(+)}_{12} &=& (U{\cal X}^nU^T + V^{\ast}{\cal Y}^{nT}V^{\dagger} - V^{\ast}{\cal U}^nU^T + U{\cal V}^nV^{\dagger})_{12}\nonumber\\
\varkappa^{n(-)}_{12} &=& (V^{\ast}{\cal X}^{n\dagger}V^{\dagger} + U{\cal Y}^{n\ast}U^T - V^{\ast}{\cal U}^{n\dagger}U^T +  U{\cal V}^{n\dagger}V^{\dagger} )_{12}.
\nonumber \\
\label{Dens}
\eea
The vertices in the canonical basis read:
\bea
g^{n}_{13} &=& \sum\limits_{24}{\bar v}_{1234}\rho^{n}_{42}, \nonumber\\
\gamma^{n(+)}_{12} &=& \frac{1}{2}\sum\limits_{34} {\bar v}_{1234}\varkappa^{n(+)}_{34}, \nonumber\\
\gamma_{12}^{n(-)T} &=& \frac{1}{2}\sum\limits_{34}{\bar v}^{\ast}_{1234}\varkappa^{n(-)\ast}_{34},
\label{Vert}
\eea
where the transition densities implicitly connect two arbitrary many-body states $|i_{(n)}\rangle$ and $|f_{(n)}\rangle$ via the multi-index $n$, cf. Ref. \cite{Litvinova2022}.

%\textcolor{blue}{
\section{Expectation values and symmetry relations}
\label{AppD}
The abbreviated forms (\ref{Krijcca}) and (\ref{Krijccb}) of the (a) and (b) components of the factorized dynamical kernel, respectively, are defined via the following commutator expectation values: 
\bea
\langle [V, A_{\mu\mu'}]\rangle^{nm} = \bigl[
(\Gamma^{(11)n}_{\mu\gamma} {\cal X}^{m}_{\mu'\gamma} + \Gamma^{(20)n}_{\mu\gamma} {\cal V}^{m}_{\mu' \gamma} \nonumber\\ - \frac{1}{2} \Gamma^{(20)n}_{\mu \mu'}\delta_{i_m f_m})
%- \left(\begin{array}{c}\mu \leftrightarrow \mu'\\ n \rightarrow {\bar n} \\ m %\rightarrow {\bar m}\end{array} \right)
- (\mu \leftrightarrow \mu')
\bigr] 
= - \langle [V, A^{\dagger}_{\mu\mu'}] \rangle^{{\bar n}{\bar m}\ast} \nonumber\\
\eea
and
\bea
\langle [V, C_{\mu\mu'}] \rangle^{nm} = \bigl[
(\Gamma^{(11)n}_{\gamma\mu} {\cal U}^{m}_{\gamma\mu'} + \Gamma^{(20)n}_{\mu'\gamma} {\cal Y}^{m}_{\mu \gamma}) \nonumber \\
%- \left(\begin{array}{c}\mu \leftrightarrow \mu'\\ n \rightarrow {\bar n} \\ m %\rightarrow {\bar m}\end{array} \right)^{\ast}
- (\mu \leftrightarrow \mu')^{\dagger}
\bigr] \nonumber \\
= \bigl[
(\Gamma^{(11){\bar n}\ast}_{\mu\gamma} {\cal U}^{{\bar m}\ast}_{\mu'\gamma} - \Gamma^{(20){\bar n}\ast}_{\mu\gamma} {\cal Y}^{{\bar m}\ast}_{\mu'\gamma}) \nonumber \\
- (\mu \leftrightarrow \mu')^{\dagger}
\bigr]
= - \langle [V, C^{\dagger}_{\mu\mu'}] \rangle^{{\bar n}{\bar m}\ast},
\eea
where 
\be
(\mu \leftrightarrow \mu')^{\dagger} = \left(\begin{array}{c}\mu \leftrightarrow \mu' \\ n \leftrightarrow {\bar n} \\ m \leftrightarrow {\bar m}\end{array}\right) 
= {\cal AS}^{\dagger}.
\ee
The following symmetry relations between the transition density components were employed:
\bea
{\cal X}_{\mu\mu'}^n &\equiv& \langle i | \alpha_{\mu'}\alpha_{\mu} | f \rangle 
=  \langle f | \alpha_{\mu}^{\dagger}\alpha_{\mu'}^{\dagger} | i \rangle ^{\ast} \equiv {\cal Y}_{\mu\mu'}^{{\bar n} \ast} \\
{\cal U}_{\mu\mu'}^n &\equiv& \langle i | \alpha_{\mu}^\dagger \alpha_{\mu'} | f \rangle 
=  \langle f | \alpha_{\mu'}^{\dagger}\alpha_{\mu}| i \rangle ^{\ast} \equiv {\cal V}_{\mu\mu'}^{{\bar n} \ast} .
\eea
\section{Double commutators of the static kernel}
\label{AppE}
The diagonal matrix elements of the static kernel read:
\bea
T^{0[11]}_{\mu \mu' \nu \nu'} &=& - \langle [[V, \Psi^{[1]}_{\mu \mu'}], \Psi^{[1]\dagger}_{\nu \nu'}] \rangle
= - \langle [[V, A_{\mu \mu'}], A^{\dagger}_{\nu \nu'}] \rangle \nonumber \\
&=& - \langle [[V, \Psi^{[2]}_{\mu \mu'}], \Psi^{[2]\dagger}_{\nu \nu'}] \rangle ^{\ast}
= - \langle [[V, A^{\dagger}_{\mu \mu'}], A_{\nu \nu'}] \rangle ^{\ast} \nonumber\\
&=& T^{0[22]*}_{\mu \mu' \nu \nu'},
\label{T0sym}
\eea
%\bea
%[V, A_{\mu \mu'}] &=& %-2\Bigl[\left(H^{40}_{\rho\mu\gamma\gamma'}C_{\rho\mu'}A^{\dagger}_{\gamma\gam%ma'}-(\mu \leftrightarrow \mu') + %H^{40}_{\mu\mu'\gamma\gamma'}A^{\dagger}_{\gamma\gamma'}\right) \nonumber \\
%&&\qquad - %\left(H^{40}_{\gamma\mu\rho\rho'}A^{\dagger}_{\rho\rho'}C_{\gamma\mu'} - (\mu %\leftrightarrow \mu') + %H^{40}_{\mu\mu'\rho\rho'}A^{\dagger}_{\rho\rho'}\right)\Bigr] \nonumber \\
%&&\; + %\Bigl[H^{31}_{\rho\rho'\mu\gamma}\left(A^{\dagger}_{\rho\rho'}A_{\mu'\gamma} %- 2C^{\dagger}_{\mu'\rho}C_{\rho'\gamma} \right) + H^{31 %\ast}_{\rho\rho'\gamma\mu}A_{\mu'\gamma}A_{\rho\rho'} - (\mu \leftrightarrow %\mu')\Bigr] \nonumber \\
%&&\; + \frac{1}{4}\Bigl[\left([H^{22}_{\rho\mu\gamma\gamma'}C_{\rho\mu'} - %(\mu \leftrightarrow \mu') + H^{22}_{\mu\mu'\gamma\gamma'}]A_{\gamma\gamma'} %\right)
%- \left( \begin{array}{c}
%\gamma \leftrightarrow \rho \\
%\gamma' \leftrightarrow \rho'
%\end{array} \right)\Bigr]
%\eea
\bea
T^{0[11]}_{\mu \mu' \nu \nu'} 
&=& - \langle [[V, A_{\mu \mu'}], A^{\dagger}_{\nu \nu'}] \rangle \nonumber \\
&=& \Biggl[
\Bigl\{
\,2\, H^{40}_{\rho \mu \gamma \gamma'} 
\bigl( 
{\cal D}^{[21]}_{\nu \rho \gamma \gamma'} 
+ 
{\cal D}^{[21]}_{\gamma \gamma' \nu \rho}
\bigr)
\delta_{\nu' \mu'} \nonumber \\
%&&\quad 
&+& 
H^{31}_{\rho \rho' \mu \gamma}
\Bigl(
\bigl(
{\cal D}^{[24]}_{\rho \rho' \nu' \gamma}\,\delta_{\mu' \nu}
- (\mu' \leftrightarrow \gamma)
\bigr) \nonumber\\
&-& 2\bigl(
{\cal D}^{[24]}_{\nu \rho \rho' \gamma}\delta_{\nu' \mu'}
+ 
{\cal D}^{[41]}_{\mu' \rho \nu \rho'}\,\delta_{\nu' \gamma}
\bigr)
\Bigr) \nonumber \\
%&&\quad +\,
&+& H^{31 *}_{\rho \rho' \gamma \mu}
\Bigl(
\bigl(
{\cal D}^{[32]}_{\nu' \gamma \rho \rho'}\,\delta_{\mu' \nu}
- (\mu' \leftrightarrow \gamma)
\bigr) \nonumber\\
&+&
\bigl(
{\cal D}^{[14]}_{\mu' \gamma \nu' \rho'}\,\delta_{\rho \nu}
- (\rho \leftrightarrow \rho')
\bigr)
\Bigr) \nonumber \\
%&&\quad +\,
&+&
H^{22}_{\gamma \mu \rho \rho'}
\Bigl(
{\cal D}^{[34]}_{\gamma \mu' \nu' \rho'}\,\delta_{\rho \nu}
- (\rho \leftrightarrow \rho')
\Bigr) \nonumber \\
&+& 
H^{22}_{\gamma \mu \rho \rho'}\,
{\cal D}^{[22]}_{\nu \gamma \rho \rho'}\,\delta_{\nu' \mu'}
\Bigr\}
- \{ \mu \leftrightarrow \mu' \}
\Biggr] \nonumber\\
&-& [ \nu \leftrightarrow \nu' ],
\eea
\bea
T^{0[33]}_{\mu \mu' \nu \nu'} &=& - \langle [[V, \Psi^{[3]}_{\mu \mu'}], \Psi^{[3]\dagger}_{\nu \nu'}] \rangle
= - \langle [[V, C_{\mu \mu'}], C^{\dagger}_{\nu \nu'}] \rangle \nonumber \\
&=& - \langle [[V, \Psi^{[4]}_{\mu \mu'}], \Psi^{[4]\dagger}_{\nu \nu'}] \rangle ^{\ast}
= - \langle [[V, C^{\dagger}_{\mu \mu'}], C_{\nu \nu'}] \rangle ^{\ast} \nonumber\\
&=& T^{0[44]*}_{\mu \mu' \nu \nu'},
\eea
\bea
T^{0[33]}_{\mu\mu'\nu\nu'} &=& - \langle [[V, C_{\mu\mu'}], C^{\dagger}_{\nu\nu'}] \rangle \nonumber \\
&=& \Bigl[4H^{40}_{\rho\rho'\gamma\mu'}\Big( {\cal D}^{[21]}_{\rho\nu'\gamma\mu}\,\delta_{\rho'\nu} - {\cal D}^{[21]}_{\rho'\nu'\gamma\mu}\,\delta_{\rho\nu} \nonumber \\
&&\; + {\cal D}^{[21]}_{\rho\rho'\gamma\nu'}\,\delta_{\mu\nu} - {\cal D}^{[21]}_{\rho\rho'\mu\nu'}\,\delta_{\gamma\nu} \Big) \nonumber \\
&&\; - H^{31}_{\rho\rho'\gamma\mu}\Big( {\cal D}^{[24]}_{\rho'\nu'\gamma\mu'}\,\delta_{\rho\nu} - {\cal D}^{[24]}_{\rho\nu'\gamma\mu'}\,\delta_{\rho'\nu} \nonumber \\
&&\; + {\cal D}^{[24]}_{\rho\rho'\gamma\nu}\,\delta_{\mu'\nu'} - {\cal D}^{[23]}_{\rho\rho'\mu'\nu'}\,\delta_{\gamma\nu} \Big) \nonumber \\
&&\; + H^{31}_{\rho\rho'\mu'\gamma}\Big( {\cal D}^{[23]}_{\rho'\nu'\gamma\mu}\,\delta_{\rho\nu} - {\cal D}^{[23]}_{\rho\nu'\gamma\mu}\,\delta_{\rho'\nu} \nonumber \\
&&\; + {\cal D}^{[24]}_{\rho\rho'\mu\nu}\,\delta_{\gamma\nu'} - {\cal D}^{[23]}_{\rho\rho'\gamma\nu'}\,\delta_{\mu\nu} \Big) \nonumber \\
&&\; - 2H^{31}_{\gamma\mu'\rho\rho'}\Big( {\cal D}^{[24]}_{\mu\nu'\rho\rho'}\,\delta_{\gamma\nu} - {\cal D}^{[24]}_{\gamma\nu'\rho\rho'}\,\delta_{\mu\nu} \nonumber \\
&&\; + {\cal D}^{[24]}_{\gamma\mu\rho\nu}\,\delta_{\rho'\nu'} - {\cal D}^{[23]}_{\gamma\mu\rho'\nu'}\,\delta_{\rho\nu} \Big) \nonumber \\
&&\; - \frac{1}{2}\, H^{22}_{\gamma\mu'\rho\rho'}\Big( {\cal D}^{[22]}_{\mu\nu'\rho\rho'}\,\delta_{\gamma\nu} - {\cal D}^{[22]}_{\gamma\nu'\rho\rho'}\,\delta_{\mu\nu} \nonumber \\
&&\; + {\cal D}^{[22]}_{\gamma\mu\rho\nu}\,\delta_{\rho'\nu'} - {\cal D}^{[22]}_{\gamma\mu\rho'\nu}\,\delta_{\rho\nu'} \Big)\Bigr] \nonumber \\
&&\; - 
%{\cal AS}^{\ast}
\Bigl[ \begin{array}{c}
   \mu \leftrightarrow \mu' \\
   \nu \leftrightarrow \nu'
\end{array}\Bigr]^{\ast},
\label{T033}
\eea
\iffalse
\bea
{\cal D}_{\mu \mu' \nu \nu'} &=& \sum\limits_{n}p_{i_n}
\left(\begin{array}{cccc}
    {\cal Y}^{n*}_{\mu \mu'} {\cal Y}^{n}_{\nu \nu'} &
    {\cal Y}^{n*}_{\mu \mu'} {\cal X}^{n}_{\nu \nu'} &
    {\cal Y}^{n*}_{\mu \mu'} {\cal V}^{n}_{\nu \nu'} &
    {\cal Y}^{n*}_{\mu \mu'} {\cal U}^{n}_{\nu \nu'} \\
    {\cal X}^{n*}_{\mu \mu'} {\cal Y}^{n}_{\nu \nu'} &
    {\cal X}^{n*}_{\mu \mu'} {\cal X}^{n}_{\nu \nu'} &
    {\cal X}^{n*}_{\mu \mu'} {\cal V}^{n}_{\nu \nu'} &
    {\cal X}^{n*}_{\mu \mu'} {\cal U}^{n}_{\nu \nu'} \\
    {\cal V}^{n*}_{\mu \mu'} {\cal Y}^{n}_{\nu \nu'} &
    {\cal V}^{n*}_{\mu \mu'} {\cal X}^{n}_{\nu \nu'} &
    {\cal V}^{n*}_{\mu \mu'} {\cal V}^{n}_{\nu \nu'} &
    {\cal V}^{n*}_{\mu \mu'} {\cal U}^{n}_{\nu \nu'} \\
    {\cal U}^{n*}_{\mu \mu'} {\cal Y}^{n}_{\nu \nu'} &
    {\cal U}^{n*}_{\mu \mu'} {\cal X}^{n}_{\nu \nu'} &
    {\cal U}^{n*}_{\mu \mu'} {\cal V}^{n}_{\nu \nu'} &
    {\cal U}^{n*}_{\mu \mu'} {\cal U}^{n}_{\nu \nu'}
\end{array} \right) \nonumber \\
&=& \sum\limits_{n}p_{i_n}
\left(\begin{array}{c}
{\cal Y}^{n*}_{\mu \mu'} \\
{\cal X}^{n*}_{\mu \mu'} \\
{\cal V}^{n*}_{\mu \mu'} \\
{\cal U}^{n*}_{\mu \mu'}
\end{array}\right)
\times
\left(\begin{array}{cccc}
{\cal Y}^{n}_{\nu \nu'} &
{\cal X}^{n}_{\nu \nu'} &
{\cal V}^{n}_{\nu \nu'} &
{\cal U}^{n}_{\nu \nu'}
\end{array}\right),
\eea
\fi
where summation is implied over the repeated indices, and the generalized superfluid two-body density is used:
\bea
{\cal D}_{\mu \mu' \nu \nu'} 
&=&
\left\langle
\left(
\begin{array}{cccc}
A_{\mu \mu'} A^{\dagger}_{\nu \nu'} &
A_{\mu \mu'} A_{\nu \nu'} &
A_{\mu \mu'} C^{\dagger}_{\nu \nu'} &
A_{\mu \mu'} C_{\nu \nu'} \\[4pt]
A^{\dagger}_{\mu \mu'} A^{\dagger}_{\nu \nu'} &
A^{\dagger}_{\mu \mu'} A_{\nu \nu'} &
A^{\dagger}_{\mu \mu'} C^{\dagger}_{\nu \nu'} &
A^{\dagger}_{\mu \mu'} C_{\nu \nu'} \\[4pt]
C_{\mu \mu'} A^{\dagger}_{\nu \nu'} &
C_{\mu \mu'} A_{\nu \nu'} &
C_{\mu \mu'} C^{\dagger}_{\nu \nu'} &
C_{\mu \mu'} C_{\nu \nu'} \\[4pt]
C^{\dagger}_{\mu \mu'} A^{\dagger}_{\nu \nu'} &
C^{\dagger}_{\mu \mu'} A_{\nu \nu'} &
C^{\dagger}_{\mu \mu'} C^{\dagger}_{\nu \nu'} &
C^{\dagger}_{\mu \mu'} C_{\nu \nu'}
\end{array}
\right)
\right\rangle \nonumber \\
&=&
\left\langle
\left(
\begin{array}{c}
A_{\mu \mu'} \\[2pt]
A^{\dagger}_{\mu \mu'} \\[2pt]
C_{\mu \mu'} \\[2pt]
C^{\dagger}_{\mu \mu'}
\end{array}
\right)
\!\!
\left(
\begin{array}{cccc}
A^{\dagger}_{\nu \nu'} &
A_{\nu \nu'} &
C^{\dagger}_{\nu \nu'} &
C_{\nu \nu'}
\end{array}
\right)
\right\rangle .
\eea
%
%%%%%%%%%%%%
%===============================================================================
\\
\\
\\
\\
\\
\bibliography{Bibliography_Jun2024}
\end{document}